\documentclass[aps, twocolumn, floatfix, superscriptaddress]{revtex4-2}

\usepackage{amscd}
\usepackage{amsmath}
\usepackage{amssymb}
\usepackage{graphicx}%
\usepackage{dsfont}
\usepackage{bm}

\usepackage{placeins}
\usepackage{multirow}

\setcitestyle{super} 

\newtheorem{rem}{Remark}[section]
\def\br{\begin{rem}}
\def\er{\end{rem}}

\def\bi{\begin{itemize}}
\def\ei{\end{itemize}}

\makeatletter
\renewcommand{\section}{\@startsection{section}{1}{\z@}%
    {-3.5ex \@plus -1ex \@minus -.2ex}
    {1.5ex \@plus .2ex}
    {\normalfont\large\bfseries\raggedright}}

\renewcommand{\subsection}{\@startsection{subsection}{2}{\z@}%
    {-2.5ex \@plus -1ex \@minus -.2ex}
    {1ex \@plus .2ex}
    {\normalfont\normalsize\bfseries\raggedright}}
\makeatother

\usepackage{newtxtext}
\usepackage{booktabs,caption} 
\usepackage{wrapfig,overpic,setspace}

\usepackage[colorlinks=true, pdfstartview=FitV, linkcolor=blue,
            citecolor=blue, urlcolor=blue]{hyperref} 
\usepackage{subcaption}
\usepackage{sidecap}
\usepackage{xcolor}

\usepackage{scalerel}

\definecolor{rred}{rgb}{0.7,0,0.1}
\definecolor{ccyan}{rgb}{0,.5,1}
\definecolor{greenrb}{rgb}{0.2,0.6,0.2}

\def\bea{\begin{equation} \begin{aligned}}
\def\eea{\end{aligned} \end{equation}}
\def\beas{\begin{equation*} \begin{aligned}}
\def\eeas{\end{aligned} \end{equation*}}
\def\bes{\begin{equation*}}
\def\ees{\end{equation*}}

\def\be{\begin{equation}}
\def\ee{\end{equation}}
\def\adots{
  \mathinner{\mkern1mu\raise1pt\hbox{.}\mkern2mu\raise4pt\hbox{.}
  \mkern2mu\raise7pt\vbox{\kern7pt\hbox{.}}\mkern1mu}}

\begin{document}

\title{Resolving Emergent Beat Patterns Through Hybrid Bayesian Learning of\protect\\
 Multilayered Stochastic Hierarchical Delay Models}

\author{Micka{\"e}l D. Chekroun}
\email{mchekroun@atmos.ucla.edu}
\affiliation{Department of Atmospheric and Oceanic Sciences, University of California, Los Angeles, CA 90095-1565, USA}
\affiliation{Department of Earth and Planetary Sciences, Weizmann Institute of Science, Rehovot 76100, Israel}

\author{Tom Dror}
\affiliation{Cooperative Institute for Research in Environmental Sciences (CIRES), University of Colorado, Boulder, CO, USA}
\affiliation{Chemical Sciences Laboratory, National Oceanic and Atmospheric Administration (NOAA), Boulder, CO, USA}

\author{Ilan Koren}
\affiliation{Department of Earth and Planetary Sciences, Weizmann Institute of Science, Rehovot 76100, Israel}

\author{Honghu Liu}
\affiliation{Department of Mathematics, Virginia Tech, Blacksburg, VA 24061, USA}

\begin{abstract}
\noindent{\bf ABSTRACT}\protect\\
 Modeling emergent, multiscale patterns in complex systems remains a persistent interdisciplinary  challenge, particularly when deciphering transient, highly coupled dynamics from severely limited data. To overcome the inherent spectral sparsity of standard finite-dimensional stochastic models, we introduce Multilayered Stochastic Hierarchical Delay Models (MSHDMs). By embedding dynamics within a hierarchical delay structure, MSHDMs leverage an infinite-dimensional phase space to engineer the high spectral density required for generating complex Amplitude-Frequency Modulation (AM-FM) dynamics, avoiding data-heavy neural network parameterizations. To reliably calibrate these sensitive structures from short observational records, we deploy a hybrid offline-online Bayesian optimization framework. Bypassing the failure points of traditional trajectory matching, our algorithm autonomously learns optimal latent coordinates and continuous fractional delays by strictly enforcing spectral consistency against the empirical Global Wavelet Spectrum. Applying this methodology to high-resolution satellite observations of continental cloud fields, the resulting stochastic emulator captures the full spatiotemporal coherence of the turbulent system using just 6 hyperparameters. The wavelet scalograms demonstrate how the model natively generates emergent wave-packet dynamics and cross-scale energy cascades, recovering semidiurnal, mesoscale, and individual cloud timescales. Supported by rigorous mathematical foundations and robust data-driven calibrations, MSHDMs thus provide a highly compressible, general-purpose tool for resolving latent AM-FM beat patterns across diverse disciplines.
\end{abstract}

\date{\today}%

\maketitle


\section*{Introduction}

\noindent Modeling the dynamics of complex systems is fundamentally challenged by the presence of multiscale interactions that manifest as transient, emergent patterns. Across diverse disciplines---ranging from climate dynamics and fluid turbulence to nonlinear optics and biological feedback networks---systems characterized by a multiplicity of interacting timescales routinely produce complex temporal signatures \cite{buzsaki2004neuronal,buzsaki2012mechanisms, soriano2013complex,ghil2008climate,LucariniChekroun2023}. A hallmark of such multiscale coupling is the emergence of beat patterns, mathematically described as coupled Amplitude and Frequency Modulation (AM-FM) \cite{hoppensteadt1999oscillatory,tort2010measuring,ghil2002advanced,barnes2024phase}. Accurately recovering and simulating these macroscopic organizing features from limited, high-dimensional observational data remains a persistent interdisciplinary hurdle.

To faithfully simulate such systems, the parameterization of neglected and unresolved subgrid variables is paramount. These parameterizations are typically sought to provide closure models---or reduced-order models---that emulate the fully resolved problem while solving only a subset of variables of interest \cite{Pavliotis2008, E.Lu.2011, abdulle2012heterogeneous}. In a modern data-driven context, this challenge often involves discovering latent variables that are not simply a subset of the original state, but are new, simpler-to-learn coordinates capturing the essential dynamics. While modern machine learning excels at identifying these low-dimensional representations, models that rely purely on deep recurrent neural networks (RNNs) often suffer from a critical lack of interpretability, complicating the extraction of new physical insights. 

This difficulty underscores why traditional closure formalisms remain crucial: they provide the theoretical guidance required for stable, accurate, and physically interpretable dynamics. Such formalisms draw from a rich heritage, including turbulence theory \cite{kraichnan1987eddy, kraichnan1989there}, quantum field theories \cite{collins1985renormalization}, and the stochastic modeling of unresolved space-time scales in climate models \cite{hasselmann1976stochastic, majda2001mathematical, LucariniChekroun2023}. Foremost among these is the Mori-Zwanzig (MZ) framework from statistical physics \cite{zwanzig_memory_1961, zwanzig2001nonequilibrium,mori_transport_1965, Chorin_al02, GKS04, Chorin_Hald-book}, which rigorously dictates that projecting a high-dimensional system onto a reduced set of resolved variables yields a Generalized Langevin Equation, requiring  the proper resolution of Markovian terms  \cite{CLM20_closure} (i.e., conditional expectation),  non-Markovian memory convolution integrals, and stochastic noise \cite{zhu2018estimation,chekroun2021stochastic} to account for unresolved subgrid/unobserved processes; see e.g.~\cite{parish2017dynamic, brennan2018data, santos2021reduced, tian2021data, LucariniChekroun2023} and references therein.

Recently, the MZ framework has provided a powerful theoretical backbone for data-driven closures acting directly on learned embeddings \cite{lin2021data,lin2021data_Koopman,lin2023regression}. Seeking to bridge the gap between deep learning and physically interpretable dynamics, recent architectures deploy nonlinear autoencoders to lift complex systems into invariant Koopman latent spaces, subsequently relying on specialized RNNs or Long Short-Term Memory (LSTM) networks to parameterize the MZ non-Markovian memory integrals \cite{gupta2021neural,qi2022machine,menier2025interpretable, gupta2025mori}. While these MZ-inspired architectures offer a leap in interpretability compared to pure black-box models, their reliance on highly parameterized neural networks to learn the memory closures introduces significant computational complexity. Forcing an RNN to learn dense, non-stationary frequency wandering requires intensive data footprints, potentially subject to severe overfitting when trained on the short, transient datasets typical of certain real-world multiscale phenomena such as diurnal cloud field's organizations considered in this study.

The theoretical justification for utilizing delay equations as effective reduced-order models rests on a rich mathematical foundation independently of the MZ formalism. It is well established that Delay Differential Equations (DDEs) can be rigorously derived from specific classes of Partial Differential Equations (PDEs) that support wave propagation, as detailed in Hale's seminal work \cite{Hale_Lunel} and documented through various examples. These include transport phenomena such as the telegraph equations, as well as delayed-oscillator models derived from El Niño-Southern Oscillation (ENSO) PDE models built upon the shallow-water equations \cite{Neelin_al98,Galanti_al00, chekroun_glatt-holtz,falkena2019derivation}. When viewed through the lens of the MZ projection formalism, the emergence of memory effects can manifest in multiple mathematical architectures. In many complex interacting systems, the MZ memory integral necessitates continuous convolution kernels (e.g., Gamma kernels), which can be elegantly resolved by recasting the integro-differential equations into a Markovian framework within an extended phase space of hidden auxiliary variables \cite{Majda_Harlim2012, MSM2015, santos2021reduced, LucariniChekroun2023}. Yet, for other applications, discrete delays may provide an appealing alternative route to approximate MZ memory terms—and thus the effects of the unresolved dynamics—as they offer a more direct physical interpretation in terms of distinct propagation times \cite{falkena2019derivation}. However, distinguishing whether the underlying physical memory is best represented by discrete or distributed delays remains inherently ambiguous when approached from a strictly data-driven perspective.

Within this context, a growing body of recent work has actively pursued the data-driven discovery of DDEs directly from time-series data. Extensions of the Sparse Identification of Nonlinear Dynamics (SINDy) framework, for example, have successfully incorporated discrete delays using augmented libraries \cite{sandoz2023sindy} and Bayesian optimization to identify optimal lag times \cite{pecile2025data}. Similarly, deep learning approaches have introduced Neural Delay Differential Equations, leveraging adjoint methods to parameterize non-Markovian closures \cite{monsel2024neural}. While these advancements highlight a strong momentum toward delay-based modeling, their methodological architectures are fundamentally designed to isolate explicit, deterministic nonlinear interactions and specific discrete lag times. Frameworks grounded in pointwise derivative fitting (such as SINDy) excel at identifying these explicit governing functional forms.

However, their primary structural objective is not the native generation of highly coupled, multiscale AM-FM beat patterns that characterize the spontaneous emergence of wave packets encountered across a wealth of disciplines. Such phenomena range from biological systems---such as cell membrane dynamics \cite{al2018multifrequency} and oscillating neuronal circuits \cite{buzsaki2004neuronal, english2017pyramidal}---to engineered communication systems \cite{maragos1993energy} and astrophysical gravitational waves \cite{abbott2016observation}. These beat patterns are also commonly encountered in multiscale atmospheric flows \cite{plougonven2007inertia, vanneste2013balance, CLM17_Lorenz9D, LucariniChekroun2023}, where bursts of inertia-gravity waves are known to challenge standard neural parameterizations \cite{chekroun2024high}.

This leaves a critical modeling framework gap. Standard finite-dimensional stochastic models (SDEs)---even those operating in algorithmically learned latent spaces---are inherently limited by a sparse spectral resolution. Forcing such models to generate the dense, highly clustered frequency geometries required for complex AM-FM beat patterns necessitates a massive expansion of the latent space, particularly as the number of interacting frequencies grows. To overcome this structural limitation without resorting to data-hungry neural network closures, we introduce the parsimonious class of \textit{Multilayered Stochastic Hierarchical Delay Models (MSHDMs)}. These are formulated as stochastic systems of coupled linear DDEs with hierarchical delays \cite{ruschel2021spectrum}, where individual subsystems explicitly correspond to distinct groups with disparate timescales.

The efficacy of this approach relies on a fundamental, yet historically overlooked, mathematical property: despite their apparent structural simplicity, linear DDEs possess an infinite-dimensional phase space capable of immense temporal complexity. By enforcing a strict scale separation between these delays ($\tau_1 \ll \tau_2 \ll \dots$), such models natively engineer dense spectral geometry---characterized by branches of nearly degenerate, marginally stable eigenvalues \cite{ruschel2021spectrum,lichtner2011spectrum}---that finite-dimensional SDEs struggle to capture. As we demonstrate in this study, it is precisely the continuous, pairwise interaction of these densely packed modes that serves as the mathematical engine for emergent beat patterns. Consequently, the MSHDM provides a robust, highly compressible closure framework that is structurally predisposed to generate coupled AM-FM dynamics, unlocking the ability to emulate latent wave-packet phenomena even from severely limited observational records.

Deploying such a sensitive, delay-driven architecture on real-world data, however, requires a paradigm shift in inverse stochastic modeling. Because AM-FM dynamics are characterized by wandering frequencies and transient, nonlinear phase shifts, traditional regression techniques based on point-by-point trajectory matching frequently fail, blinding the optimization algorithm to models that are potentially dynamically and spectrally accurate. To solve this, we propose a hybrid offline-online calibration strategy embedded within a Bayesian optimization framework. The methodology learns the structural model components---specifically, the latent reduced coordinates (derived via Partial Least Squares, PLS) and the hierarchical delays associated with distinct variability groups---by balancing deterministic precision with spectral fidelity. Within the optimization loop, the instantaneous linear coefficients and stochastic noise covariance are evaluated ``offline'' via regularized Ordinary Least Squares (OLS) on the in-sample training window. Concurrently, the delays and PLS coefficients are evaluated ``online'' by simulating the out-of-sample stochastic trajectory and strictly penalizing deviations in the empirical spectral geometry using the Global Wavelet Spectrum (GWS) and when possible, spectral kurtosis.  Wavelet-based approaches have been advocated to measure phase coherence and frequency modulation \cite{barnes2024phase}.  Our  hybrid offline-online philosophy enhances the resulting emulator's ability to capture the broadband frequency distribution and intermittent bursts contained in the observed dataset.

To demonstrate the full force of this hybrid offline-online Bayesian framework, we require a rigorous physical application. Unlike the aforementioned ENSO applications, which have benefited from an extensive historical lineage, conceptual delay models remain in their infancy within cloud physics \cite{koren2011aerosol, koren2017exploring, chekroun2020efficient,Chekroun_al22SciAdv}. More broadly, the robust data-driven inverse modeling of dynamic cloud decks remains a largely unsolved challenge, leaving many complex cloud systems without even a basic conceptual deciphering. Advancing data-driven delay models in this domain could offer critical new insights into cloud fields that remain notoriously resistant to representation in global climate models. This is particularly true for continental shallow cumulus clouds, which play a crucial role in Earth's radiation budget \cite{dror2022uncovering,liu2023opposing}, yet whose mesoscale organization---such as the distinctive grid shapes found in the ``Green-Cumulus'' (GreenCu) subset \cite{Dror2021}---remains systematically unresolved by standard numerical schemes. To confront this profound modeling gap, we apply our framework to high-resolution GOES-16 satellite observations of these GreenCu fields, presenting a highly complex inverse modeling target characterized by turbulent, non-stationary wave packets and deeply coupled, multiscale dynamics.

Beyond their climatological importance, we select these notoriously challenging cloud fields \cite{nair_1998, heus_2013, dror_2020} as our primary application of our MSHDM framework because the evolution of these cloud fields represents a quintessential, real-world physical manifestation of transient AM-FM dynamics. Tied to the diurnal cycle, the mesoscale organization is governed by convective Inertia-Gravity Waves (IGWs) that propagate and ``puff'' in the form of  localized wave packets---a literal emergent beat pattern \cite{Dror2021}. Because these phenomena are transient and short-lived, long-term forecasting is ill-posed. Instead, we utilize the MSHDM to construct a ``Groundhog Day'' stochastic emulator. By training on a short, high-resolution subset of GOES-16 satellite principal components, the emulator generates massive synthetic ensembles of the observed diurnal event, allowing us, in particular, to uncover the latent structural features and cross-scale energy cascades driving the cloud deck.

The remainder of this paper is organized as follows. First, we present the foundational theory of Multilayered Stochastic Hierarchical Delay Models, detailing the spectral analysis of hierarchical delays and deriving the general $M$-interacting modes AM-FM equation. We then demonstrate our Inverse Stochastic Modeling methodology, utilizing a Bayesian optimization approach to recover the AM-FM dynamics of a theoretical scalar DDE toy model based strictly on spectral matching. Finally, we deploy the full force of the calibrated MSHDM framework onto the GOES-16 cloud dataset. We perform an $L_1$-norm quaternary decomposition of the model's PLS coeffcients to extract the physical drivers bridging the larger, slow scales and the smaller, fast scales, and demonstrate the framework's ability to seamlessly generate the complex out-of-sample wave packet dynamics governing convective cloud field organization.


\section*{Results}
\subsection*{Multilayered Stochastic Hierarchical Delay Models}

To overcome the inherent spectral sparsity of standard finite-dimensional stochastic systems, this study introduces the class of Multilayered Stochastic Hierarchical Delay Models (MSHDMs). 
We define the MSHDM framework as a stochastic system of coupled linear DDEs partitioned across $L$ interacting spectral or topological layers (e.g., macroscopic forcing, mesoscale organization, and small-scale turbulence). 

Let $\mathbf{x} = [\mathbf{x}_1^T, \mathbf{x}_2^T, \dots, \mathbf{x}_L^T]^T \in \mathbb{R}^d$ represent the full multiscale state vector. Here, each subvector $\mathbf{x}_\ell \in \mathbb{R}^{d_\ell}$ isolates the dynamics of the $\ell$-th layer, corresponding to a distinct grouping of characteristic timescales for the system at hand (typically ordered from the slowest to the fastest). The system is governed by the coupled layer-wise equations:
\bea\label{Eq_MSHDM_General}
\mathrm{d}\mathbf{x}_\ell(t) &= \bigg[ \sum_{j=1}^L \mathbf{A}_0^{\ell j} \mathbf{x}_j(t) + \sum_{j=1}^L \sum_{k=1}^{p_\ell} \mathbf{A}_k^{\ell j} \mathbf{x}_j(t-\tau_k^{\ell}) \bigg] \mathrm{d}t \\
& \hspace{2cm}+\mathbf{\Sigma}_\ell \mathrm{d}\mathbf{W}_{\ell}(t),  \;\text{ for } \ell = 1, \dots, L,  \\
&\mbox{with }  0 \leq  \tau_1^{\ell} \ll \tau_2^{\ell} \ll \cdots \ll \tau_{p_\ell}^{\ell},
\eea
where $\mathbf{A}_0^{\ell j}$ governs the instantaneous interactions from layer $j$ to layer $\ell$, and the delayed matrices $\mathbf{A}_k^{\ell j}$ explicitly encode the layered, cross-scale temporal couplings (such as forward energy cascades and high-frequency turbulent backscatter). The term $\mathrm{d}\mathbf{W}_{\ell}(t)$ denotes a $d_\ell$-dimensional vector of mutually independent Wiener processes, and $\mathbf{\Sigma}_\ell$ is a positive-definite noise covariance matrix specific to layer $\ell$. In this work, the notation $\ll$ means either large (absolute) gaps between the delays within a layer as in \cite{ruschel2021spectrum}, or delays covering disparate timescales from a physical viewpoint (e.g., minutes vs. hours).

While nonlinear components could naturally be included in such a model formulation, we restrict our theoretical focus to the linear case for analytic purposes. As demonstrated below, the interaction of hierarchical delays is mathematically sufficient to generate dense spectral geometries capable of supporting the complex AM-FM behaviors observed in empirical data.

To gain understanding, we analyze a scalar MSHDM in the complex domain.
A defining characteristic of MSHDMs is their incorporation of hierarchical delays that scale as $\tau_k \sim \kappa_k \epsilon^{-k}$ for a small parameter $\epsilon \ll 1$. To describe the core mathematical mechanisms at the origin of AM-FM behavior, we isolate the deterministic skeleton ($\mathbf{\Sigma}=0$) of a single scalar layer in the complex domain ($\mathbb{C}$), governed by two highly disparate timescales:
\be\label{Eq_HDM}
\dot{x} = a x + b x(t-\tau_1) + c x(t-\tau_2),
\ee
where the hierarchical delays are $\tau_1 = 1/\epsilon$ and $\tau_2 = \kappa/\epsilon^2$. We assume that $b, c$, and $\kappa$ are real, while $a$ is complex, with $\Re(a) < 0$ ensuring a dissipative instantaneous baseline.


\subsection*{Spectral Analysis of Hierarchical Delay Models via Galerkin-Koornwinder Expansion}
To gain analytic insight into how this deterministic MSHDM (Eq.~\eqref{Eq_HDM}) structurally generates the high spectral density necessary for organizing AM-FM behaviors, we examine its characteristic equation:
\begin{equation}\label{Eq_chi_x}
\chi_x(\lambda) = \lambda - a - b e^{-\lambda\tau_1} - c e^{-\lambda\tau_2} = 0,
\end{equation}
where $\lambda = \sigma + i\omega$ are the characteristic roots (eigenvalues). The stability of the system's trivial steady state is determined by the real part of these roots, $\sigma$. This equation represents a singular perturbation problem due to the non-analytic nature of the exponential terms as $\epsilon \to 0$.

The presence of these two highly disparate delays ($\tau_1 \ll \tau_2$) is the mathematical engine for creating a {\it Spectral Hierarchy}, consisting of distinct, dense families of characteristic roots ($\Sigma_1$ and $\Sigma_2$) in the complex $\lambda$-plane. Typically, the instability analysis reveals that these dense  families are typically located  on a very narrow vertical strip in the left half-plane ($\sigma \approx 0$) \cite{ruschel2021spectrum}, as $\epsilon \to 0$. 

In practice, however, such asymptotic properties must be complemented by an analysis of the finite-size effects of the parameter $\epsilon$ on the spectral properties of the DDE. To gain theoretical insights in the case of delays which can be large yet constrained in size due to data limitations (from a data-driven modeling perspective), we adopt the Galerkin-Koornwinder (GK) approximation framework developed by Chekroun, Liu et al.~\cite{CGLW16}.
This framework has proved useful to derive reduced systems to rigorously study challenging nonlinear scenarios for DDEs such as the detection of saddle-node bifurcation of periodic orbits,  the computation of unstable periodic orbits (UPOs), and the approximation of homoclinic orbits \cite{chekroun2020efficient,Chekroun_Liu23Tipping}. GK approximations have also proved instrumental to the optimal control of DDEs \cite{CKL18_DDE}, and the design of stochastic perturbations to effectively  induce shear-induced chaos for DDEs \cite{Chekroun_al22SciAdv}. 

This robust framework enables us to approximate the solution to linear systems of DDEs by a finite-dimensional Galerkin-Koornwinder expansion \cite{CGLW16}, $x(t) \approx x_N(t)$. In this approximation, the system's evolution is effectively governed by the superposition of the dominant eigenmodes:
\be\label{Eq_approx_DDE}
x(t) \approx \sum_{k=1}^{M} D_k e^{\lambda_k t}, \quad \lambda_k = \sigma_k + i \omega_k,
\ee
where $M$ is the number of dominant modes, $\lambda_k$ are the eigenvalues of the GK matrix approximation $A_N$, and $D_k$ are complex coefficients determined by the initial condition \cite{chekroun2020efficient,Chekroun_Liu23Tipping}.  

This approach elegantly transforms the infinite-dimensional DDE problem into a finite-dimensional eigenvalue problem. Crucially, by expressing the complex temporal evolution as a clean, discrete superposition of modal exponentials, the GK framework provides profound analytical transparency. It allows us to bypass the cumbersome machinery of infinite-dimensional functional analysis (such as the explicit construction of dual operators or intricate functional state spaces) by projecting the DDE's history content and eigenfunctions onto a large, robust set of Koornwinder polynomials (a generalization of Legendre polynomials). This mathematically transparent reduction to a finite-dimensional algebraic form is precisely what makes our subsequent, rigorous decomposition of the AM-FM phase dynamics algebraically tractable and physically interpretable.

\begin{figure}
\vspace{-2ex}
\centering
    \includegraphics[width=1\linewidth, height=0.5\textwidth]{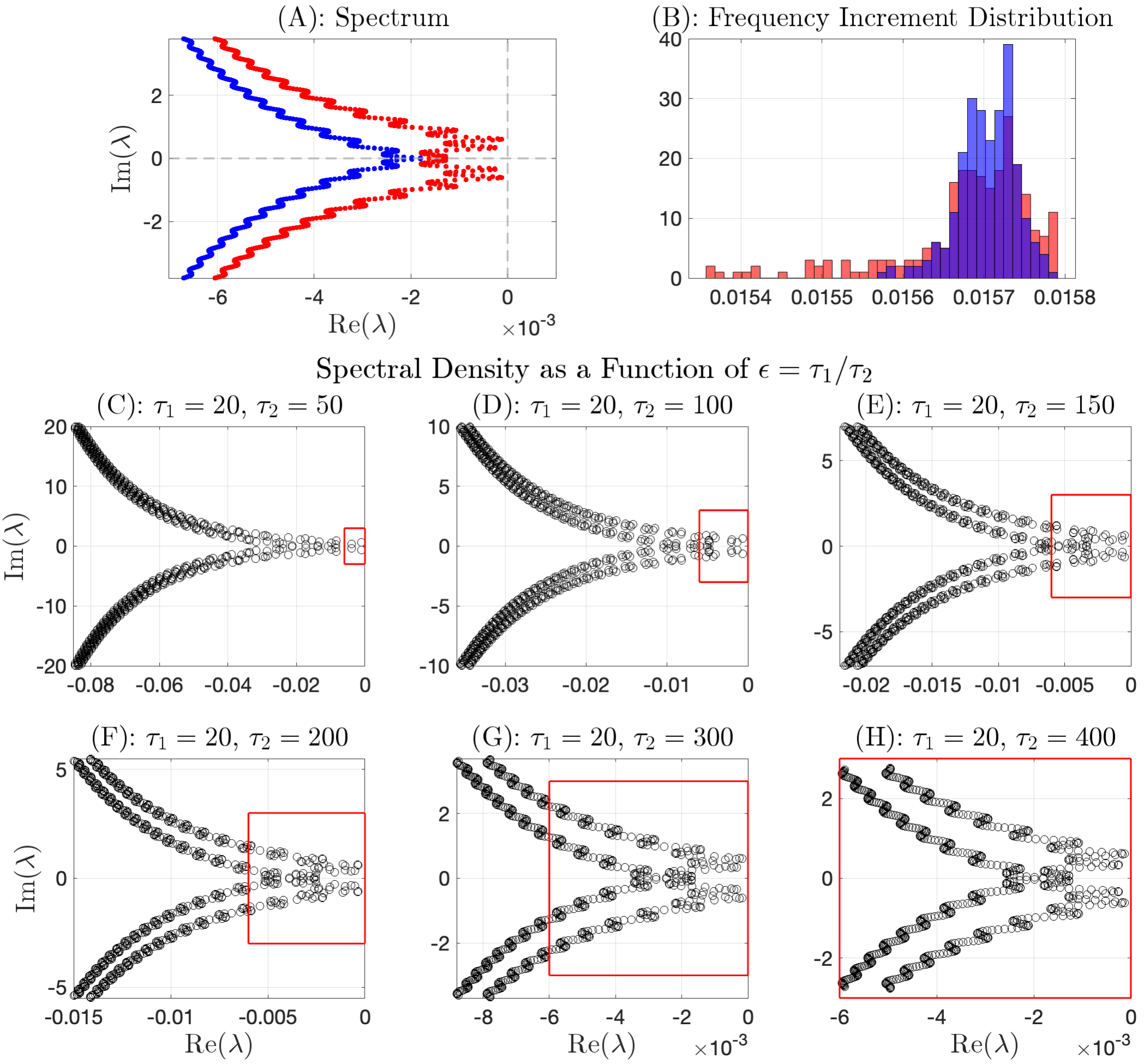}
\caption{\textbf{Spectral Density Plots.} {\bf Panel A}: Spectrum of the scalar DDE, Eq.~\eqref{Eq_HDM}, illustrating the high spectral density. {\bf Panel B}: Spectral density of the adjacent modes. The blue (resp.~red) histogram corresponds to the blue (resp.~red) branch of eigenvalues in Panel A, highlighting the extreme proximity in frequency ($\Delta\omega$). {\bf Panels C--H}: Spectrum evolution as the hierarchical delay ratio $\epsilon = \tau_1/\tau_2$ decreases from $0.4$ to $0.05$, with $\tau_1=20$ held constant. The red rectangular boxes indicate the fixed spectral region $[-0.006, 0] \times [-3, 3]$. As $\epsilon$ decreases, an increasing number of eigenvalues enter this region, resulting in a significantly denser spectral distribution. Throughout all panels, the base parameters are $a=-0.4 + 0.5i$, $b=0.1$, and $c=0.3$. Panels A and B specifically focuses on the highly hierarchical case $\epsilon=0.05$ ($\tau_2=400$).}
\label{Fig_Density}
\vspace{-.7cm}
\end{figure}

For the specific choice of $a=-0.4 + 0.5i$, $b=0.1$, $c=0.3$, and the small parameter $\epsilon=0.05$ (yielding $\tau_1=20$ and $\tau_2=400$), the GK spectrum achieves an accuracy $<10^{-5}$ for a large subset of dominant DDE eigenpairs (verified against the characteristic equation), using a truncation dimension of size $N=1600$. 

These eigenpairs form a unique spectral geometry shown in Figure \ref{Fig_Density}A ($483$ pairs are displayed). 
The spectrum is composed of two distinct, yet closely spaced, sets of eigenvalues, $\Sigma_1$ (red) and $\Sigma_2$ (blue). These modes are tightly clustered just to the left of the imaginary axis within a narrow vertical strip. Their real parts satisfy $\sigma_k \in (-\delta, 0)$, with $\delta \approx 6.5 \times 10^{-3}$. This configuration guarantees a \textit{very slow decay} ($e^{\sigma_k t}$), ensuring the modes are marginally stable and persist long enough to interact. On these branches, the eigenvalues exhibit {\it very small frequency separation}, $\Delta \omega \ll 1$, characterized by adjacent modes that are extremely close in frequency (Figure \ref{Fig_Density}B), resulting in an overall high spectral density.

Figures \ref{Fig_Density}C--H illustrate the spectrum evolution as the ratio $\epsilon$ varies from $0.4$ to $0.05$, with $\tau_1 = 20$ held constant. The red rectangular boxes highlighted in these panels indicate the fixed spectral region $[-0.006, 0] \times [-3, 3]$. It is evident that as $\epsilon$ decreases (i.e., as the delays become increasingly hierarchical), an increasing number of eigenvalues enter this region, resulting in a significantly denser spectral distribution.

To gain further insights into the generation of FM from this spectral geometry, we analyze the instantaneous frequency $\omega_{inst}(t)$ of the signal. The signal $x(t)$ is expressed in terms of its complex envelope, $A(t)$, and a carrier frequency, $\omega_c$ (e.g., $\omega_1$, or $\frac{\omega_1+\omega_2}{2}$):
\be\label{Eq_Complex_Envelope}
x(t) = e^{i\omega_c t} \underbrace{\left[ \sum_{k=1}^{M} D_k e^{\lambda_k t} e^{-i\omega_c t} \right]}_{A(t)}.
\ee
The instantaneous frequency $\omega_{inst}(t)$ is defined as the time derivative of the signal's total phase, $\Phi(t) = \omega_c t + \phi(t)$, where $\phi(t) = \arg(A(t))$ is the phase of the envelope, namely \cite[Chap.~3.3]{proakis2002solutions}:
\bes
\omega_{inst}(t) = \frac{d\Phi(t)}{dt} = \omega_c + \frac{d\phi(t)}{dt}.
\ees
The signal exhibits FM when the derivative of the envelope's phase, $\dot{\phi}(t)$, is non-constant.


The case of two interacting modes ($M=2$) is already enlightening. By exploring this scenario under the condition of near-degeneracy---where the modes are very close but not identical (echoing the DDE's dense spectrum)---we demonstrate that this property is instrumental in producing both AM and FM, as encountered in many applications; see e.g.~\cite{hoppensteadt1998thalamo,hoppensteadt1999oscillatory} and references therein. To that end, we examine the structure of the complex envelope $A(t)$:
\be\label{Eq_gene_amp_M=2}
A(t) = |D_1| e^{\sigma_1 t} e^{i[(\omega_1 - \omega_c) t + \phi_{1,0}]} + |D_2| e^{\sigma_2 t} e^{i[(\omega_2 - \omega_c) t + \phi_{2,0}]},
\ee
where $\phi_{k,0} = \arg(D_k)$. 
 
Near-degeneracy of the modes implies their frequencies and damping rates are very close to one another:
$$
\omega_1 = \omega_2 + \delta\omega \quad \text{and} \quad \sigma_1 = \sigma_2 + \delta\sigma \quad \text{with} \quad \delta\omega, \delta\sigma \ll 1.
$$
We set the initial phase to zero, $\phi_{k,0}=0$, and the center frequency to $\omega_c = \omega_2$ for simplicity. 

The complex envelope $A(t)$ becomes:
\be\label{Eq_envelop}
A(t) = |D_1|e^{(\sigma_2 + \delta\sigma) t} e^{i\delta\omega t} + |D_2|e^{\sigma_2 t}.
\ee
AM is described by the squared magnitude of the complex envelope, $|A(t)|^2$. Modulation occurs when this envelope is not constant in time ($\frac{d}{dt}|A(t)|^2 \neq 0$). We have
\beas
|A(t)|^2 = A(t) \bar{A}(t) = \bigg[&|D_1|^2 e^{2(\sigma_2 + \delta\sigma) t} + |D_2|^2 e^{2\sigma_2 t}\bigg] \\
&+ 2 |D_1||D_2| e^{(2\sigma_2 + \delta\sigma) t} \cos(\delta\omega t).
\eeas
The $|A(t)|^2$ expression clearly shows two distinct contributions that ensure AM is present whenever the modes are nearly degenerate ($\delta\omega$ and $\delta \sigma$ small). Two distinct features can then be inferred: (i) {\it Periodic Beat Modulation} (due to $\delta\omega$), and (ii) {\it Slow transient Amplitude Drift} (due to $\delta\sigma$). 

The former is generated by the term $2 |D_1||D_2| e^{(2\sigma_2 + \delta\sigma) t} \cos(\delta\omega t)$. Since the frequency difference is small ($|\delta\omega| \ll 1$), the superposition creates a {\it slow periodic beat} in the envelope magnitude oscillating at the beat frequency $\delta\omega$. This is the classic signature of AM. The slow transient amplitude drift is caused by the proximity of the damping rates ($|\delta\sigma| \ll 1$). Because the exponents $2(\sigma_2 + \delta\sigma) t$ and $2\sigma_2 t$ are close to each other, the relative contribution of Mode 1 versus Mode 2 slowly changes over time. The $e^{\delta\sigma t}$ factor in the exponent of both terms ensures that the overall envelope magnitude is slowly changing (decaying if $\sigma_{1,2}<0$). This continuous change in the relative modal weights constitutes a form of secular (long-term, non-periodic) amplitude modulation.
 
From Eq.~\eqref{Eq_gene_amp_M=2}, using the explicit expression of the envelope phase $\phi(t)=\arctan (\text{Im}(A(t))/\text{Re}(A(t)))$, and $\omega_2 =\omega_c $,  $ \phi_{k,0}=0$, we arrive, after tedious calculations, to 
\bea\label{Eq_phi_dotM=2}
\dot{\phi}(t) = &\frac{1}{|A(t)|^2}  \bigg( |D_1||D_2| e^{(2\sigma_2 + \delta\sigma) t} \bigg[ \delta\omega \cos(\delta\omega t) \nonumber \\
&\hspace{3ex}+ \delta\sigma \sin(\delta\omega t) \bigg] + |D_1|^2 e^{2(\sigma_2 + \delta\sigma) t} \delta\omega \bigg).
\eea
In a two-mode system with non-zero $\delta\omega$ and $\delta\sigma$, the amplitude modulation and phase rate modulation are inherently coupled. When the magnitude $|A(t)|$ is oscillating due to beating (AM), the instantaneous phase rate $\dot{\phi}(t)$ is also oscillating (FM).
The term:
\be\label{Eq_cross_term}
\frac{|D_1||D_2| e^{(2\sigma_2 + \delta\sigma) t}}{|A(t)|^2} \left[\delta\omega \cos(\delta\omega t) + \delta\sigma \sin(\delta\omega t)\right],
\ee
represents the periodic modulation of the instantaneous frequency. It contains both $\cos(\delta\omega t)$ and $\sin(\delta\omega t)$, meaning the instantaneous frequency oscillates at the \textit{beat frequency} $\delta\omega$. Crucially, this frequency modulation is driven by the same beat frequency that causes the AM ($\cos(\delta\omega t)$ term in $|A(t)|^2$). This is the essence of \textit{AM-FM coupling}: the periodic amplitude changes caused by the beating modes are accompanied by periodic instantaneous frequency changes. 
\textit{Periodic FM} will refer below to the presence of an oscillatory component within $\dot{\phi}(t)$, where the frequency of the oscillation is the beat frequency $\delta\omega$.

The term:
\bes
\frac{|D_1|^2 e^{2(\sigma_2 + \delta\sigma) t}}{|A(t)|^2} \delta\omega,
\ees
represents a \textit{baseline shift} in the instantaneous frequency. If $\delta\omega \neq 0$ and $|D_1| \neq 0$, the center frequency of the oscillation is shifted, primarily weighted by the magnitude of Mode 1.

The slow transient amplitude drift caused by $\delta\sigma$ influences $\dot{\phi}(t)$ in two ways. First, the overall amplitude of the periodic frequency modulation (the first term in the expression of $\dot{\phi}$) is proportional to:
$$ \frac{|D_1||D_2| e^{(2\sigma_2 + \delta\sigma) t}}{|A(t)|^2}.$$
Since both the numerator and the denominator, $|A(t)|^2$, are undergoing a slow, secular drift determined by $\delta\sigma$, the magnitude of the frequency modulation slowly changes over time.

Second, the term $\delta\sigma \sin(\delta\omega t)$ in Eq.~\eqref{Eq_cross_term} shows that the difference in damping rates, $\delta\sigma$, also contributes directly to the periodic frequency modulation at the beat frequency $\delta\omega$. This means the difference in the exponential factors not only causes amplitude drift but also actively participates in the periodic phase/frequency changes.

\begin{figure}
\vspace{-2ex}
\centering
    \includegraphics[width=1\linewidth, height=0.5\textwidth]{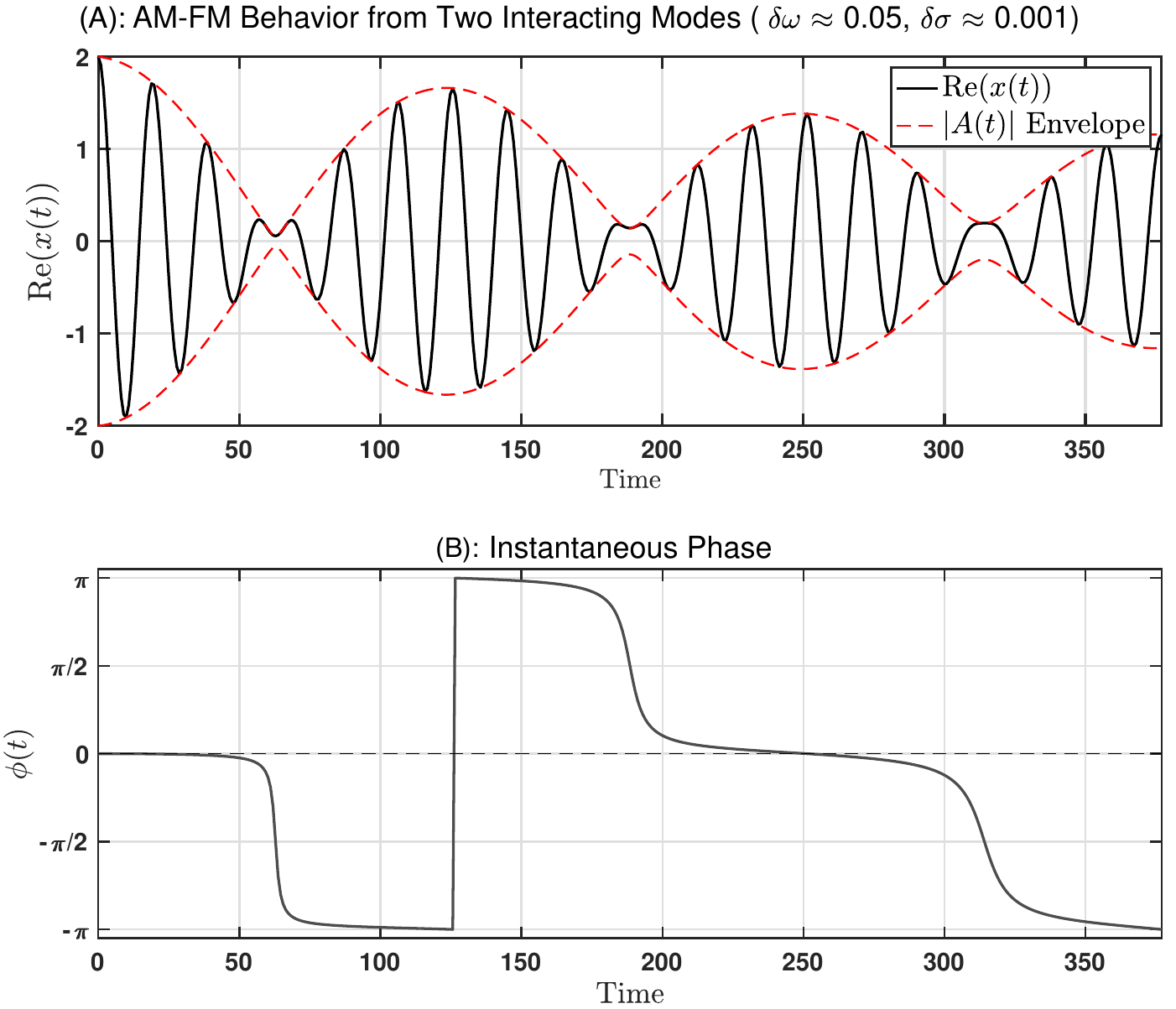}
\caption{\textbf{AM-FM Behavior from Two Interacting Modes.} Panel A shows $\textrm{Re}(x(t))$ (Eq.~\eqref{Eq_Complex_Envelope}) for $M=2$ with $\omega_c=(\omega_1+\omega_2)/2$, where $\sigma_1 = -10^{-3}$, $\omega_1=0.3$, $\sigma_2 = -2\times 10^{-3}$, $\omega_2=0.35$, $D_1=D_2=1$, and $\phi_{1,0}=\phi_{2,0}=0.$ The envelope $A(t)$ is computed from Eq.~\eqref{Eq_envelop}. Panel B shows the instantaneous phase, $\phi(t)=\arctan(\textrm{Im}(A(t))/\textrm{Re}(A(t))).$}\label{Fig_2Mode}
\vspace{-3ex}
\end{figure}

In short, in this near-degenerate system two factors control the AM-FM dynamics. Firstly, the \textit{beat frequency ($\delta\omega$)} is responsible for driving both the periodic amplitude modulation (AM) and the periodic frequency modulation (FM). Secondly, the \textit{damping difference ($\delta\sigma$)} induces the slow transient AM drift and is also instrumental in modifying the amplitude and phase of the periodic FM components. Figure \ref{Fig_2Mode} demonstrates the complex AM-FM behavior generated from two interacting modes that are nearly degenerate.  

Let us explore the limiting cases. 
In the case of a \textit{Weak Damping Difference} ($|\delta\sigma| \ll |\delta\omega|$), the system dynamics are dominated by the periodic beat generated by the frequency difference $\delta\omega$. The instantaneous phase rate, $\dot{\phi}(t)$, exhibits a strong, periodic oscillation (FM) that is strongly coupled to the AM beat ($\cos(\delta\omega t)$ term). While the small $\delta\sigma$ term is largely suppressed in the periodic modulation, its effect remains via the exponential factors, causing a slow, secular drift in the overall amplitude of the FM oscillation. When the initial amplitudes are equal ($|D_1| = |D_2|$), the beat pattern is maximized, and $\dot{\phi}(t)$ becomes a complicated function of the small parameters, showing that the instantaneous frequency swings maximally when the amplitude is minimal—a hallmark of strong AM-FM coupling. The simplified expression for $\dot{\phi}(t)$ under this equal-amplitude condition is:
\bes
\dot{\phi}(t) = \frac{e^{\delta\sigma t} \left( \delta\omega(e^{\delta\sigma t} + \cos(\delta\omega t)) + \delta\sigma \sin(\delta\omega t) \right)}{e^{2\delta\sigma t} + 1 + 2 e^{\delta\sigma t} \cos(\delta\omega t)}.
\ees
The most illuminating case is the \textit{Pure Beat Limit} ($\delta\sigma=0$ and $|D_1|=|D_2|$). Here, the long-term transient drift vanishes, leaving only the cleanest form of periodic modulation. The squared amplitude simplifies to $|A(t)|^2 = 4 D_1^2 e^{2\sigma_2 t} \cos^2\left(\delta\omega t/2\right)$, confirming pure periodic AM, with a period of the energy fluctuation equal to $2\pi/\delta \omega$. In this ideal scenario, the instantaneous phase rate becomes a constant value: $\dot{\phi}(t) = \frac{\delta\omega}{2}$. This result means the instantaneous frequency of the combined signal is stable and constant, equal to the average frequency of the two modes, $\omega_{inst} = \frac{\omega_1 + \omega_2}{2}$. Therefore, in the absence of a damping difference ($\delta\sigma$), all modulation energy is channeled into the amplitude (AM), and the frequency modulation vanishes, mathematically confirming that $\delta\sigma$ is essential for inducing periodic FM in the complex, non-ideal systems observed in nature.

\subsection*{The Case of $M$ Interacting Modes: The AM-FM Equation}

Moving beyond the two-mode approximation, the true power of the MSHDM framework lies in its dense spectral geometry consisting of $M$ interacting modes. To quantify the complex, emergent frequency modulation in this dense regime, the phase derivative, $\dot{\phi}(t)$---which dictates the instantaneous frequency shift of the envelope $A(t)$---can be elegantly expressed in terms of the complex envelope and its temporal derivative:
\be\label{Eq_phi_rate_gene}
\dot{\phi}(t)=\frac{\text{Im}(\bar{A} A')}{|A(t)|^2}.
\ee
This identity is pivotal because it allows us to rigorously evaluate the phase's rate of change in the general case, where the envelope is a vast superposition of modes (Eq.~\eqref{Eq_Complex_Envelope}):
\bes
A(t)=\sum_{k=1}^M D_k e^{\lambda_k t} e^{-i\omega_c t}.
\ees
Here, $\lambda_k = \sigma_k + i\omega_k$ are the complex eigenvalues and $D_k = |D_k| e^{i\phi_{k,0}}$ are the complex amplitude coefficients, where $\phi_{k,0}$ is the initial phase offset of the $k$-th mode at $t=0$.

The calculation of the numerator, $\text{Im}(\bar{A} A')$, involves a double summation over all $M$ modes. This double summation naturally splits into two physically distinct contributions: the diagonal terms ($k=j$), which represent the intrinsic frequency deviation of each individual mode, and the off-diagonal terms ($k \neq j$), which capture the inter-modal beating and cross-scale coupling effects.

To formalize this, we introduce $\psi_k(t)$, the phase of mode $k$ in the $\omega_c$-rotating frame, defined as $\psi_k(t) = (\omega_k - \omega_c)t + \phi_{k,0}$. The final, simplified equation for the instantaneous frequency shift is then given by:
\begin{widetext}
\vspace{-3ex}
\bea\label{Eq_dot_phi}
\dot{\phi}(t) = \frac{1}{|A(t)|^2} \sum_{k<j} |D_k||D_j| e^{(\sigma_k+\sigma_j) t} \bigg[ (\omega_k + \omega_j - 2\omega_c) \cos(\psi_k - \psi_j)& + (\sigma_k - \sigma_j) \sin(\psi_k - \psi_j) \bigg]\\
& + \frac{\sum_{k=1}^M |D_k|^2 e^{2\sigma_k t} (\omega_k - \omega_c)}{|A(t)|^2}.
\eea
\end{widetext}
See {\it Supplementary Note 1} for a full derivation.

The numerator in Eq.~\eqref{Eq_phi_rate_gene}, $\text{Im}(\bar{A} A')$, which drives the phase rate, formalizes this decomposition by splitting exactly into two structural types of terms:
\bes
\text{Im}(\bar{A} A')=\sum_{k<j} (\text{Cross-Terms})_{kj} + \sum_{k=1}^M (\text{Self-Terms})_{k}.
\ees
The architecture of this identity is profound because it perfectly separates the long-term secular drift from the transient periodic beating, providing a clear, mechanistic decomposition of the AM-FM dynamics. Our assumption that the spectrum consists of branches of nearly degenerate pairs implies that for any consecutive pair $k, j$ (where $j=k+1$), we are analyzing a very small frequency difference ($|\omega_k - \omega_j| = |\delta\omega_{kj}| \ll 1$), alongside a very small damping difference ($|\sigma_k - \sigma_j| = |\delta\sigma_{kj}| \ll 1$).

The {\it Self-Terms} form the second part of Eq.~\eqref{Eq_dot_phi}:
\bes
(\text{Self-Terms})=\sum_{k=1}^M |D_k|^2 e^{2\sigma_k t} (\omega_k - \omega_c).
\ees
This summation represents a weighted, \textit{instantaneous baseline} for the frequency modulation. Each mode $k$ contributes a constant frequency shift $(\omega_k - \omega_c)$ proportional to its instantaneous power $|D_k|^2 e^{2\sigma_k t}$. Crucially, this baseline itself is \textit{secularly modulated} (slowly drifting) due to the $\sigma_k$ decay terms. As the modes with larger (less negative) damping rates $\sigma_k$ grow in relative power, the overall instantaneous frequency baseline slowly drifts towards the $\omega_k$ of those dominant modes. This process establishes a form of \textit{long-term FM drift} strictly driven by the transient AM drift (due to $\delta\sigma$).

The {\it Cross-Terms} explicitly quantify the instantaneous frequency shift generated by inter-modal beating, driven by the coupling between the relative phases $(\psi_k - \psi_j)$. This sum is dominated by contributions from the nearly degenerate pairs. The term involving $\cos(\psi_k - \psi_j)$ oscillates at the beat frequency $\delta\omega_{kj}=(\omega_k - \omega_j)$. Its amplitude is proportional to the frequency differences relative to the carrier frequency, yielding the expression:
\bes
|D_k||D_j| e^{(\sigma_k+\sigma_j) t} \cdot (\omega_k + \omega_j - 2\omega_c)\cos(\delta\omega_{kj} t + \phi_{kj}),
\ees
where $\phi_{kj} = \phi_{k,0} - \phi_{j,0}$ is the initial phase offset of mode $k$ relative to mode $j$. Notably, this recovers the two-mode result when $\phi_{k,0}=0$ and $\omega_c=\omega_2$, as the coefficient collapses down to $\delta\omega$. 

This result establishes a fundamental dynamical link: because the overall envelope power $|A(t)|^2$ also contains terms oscillating at the beat frequency $\delta\omega_{kj}$ (the AM beat), the appearance of this exact frequency in the phase derivative confirms that \textit{periodic AM is intrinsically accompanied by periodic FM} for every interacting mode pair $(k, j)$.

Furthermore, the term involving $\sin(\psi_k - \psi_j)$ is proportional to the \textit{damping difference} $(\sigma_k - \sigma_j)=\delta\sigma_{kj}$:
$$
|D_k||D_j| e^{(\sigma_k+\sigma_j) t} \cdot \delta\sigma_{kj} \sin(\delta\omega_{kj} t + \phi_{kj}).
$$
The difference in damping rates, $\delta\sigma_{kj}$, does not merely drive the long-term AM drift; it introduces an \textit{additional periodic FM component} oscillating at the beat frequency $\delta\omega_{kj}$, but shifted $90^\circ$ out of phase with the primary beat. This quadrature interaction causes a complex periodic fluctuation in the instantaneous phase rate (FM), establishing that non-uniform dissipation across the spectrum is mathematically essential for generating wandering frequencies.

Ultimately, the general $M$-mode equation demonstrates that AM-FM coupling is fundamentally a non-trivial superposition of pairwise interactions. Instead of a single beat frequency, the system exhibits \textit{Multiple Beats}: a dense collection of periodic interactions ($\delta\omega_{kj}$) simultaneously driving the envelope power ($|A(t)|^2$) and the instantaneous phase rate ($\dot{\phi}(t)$). 

\begin{figure}
\vspace{-2ex}
\centering
    \includegraphics[width=1\linewidth, height=0.5\textwidth]{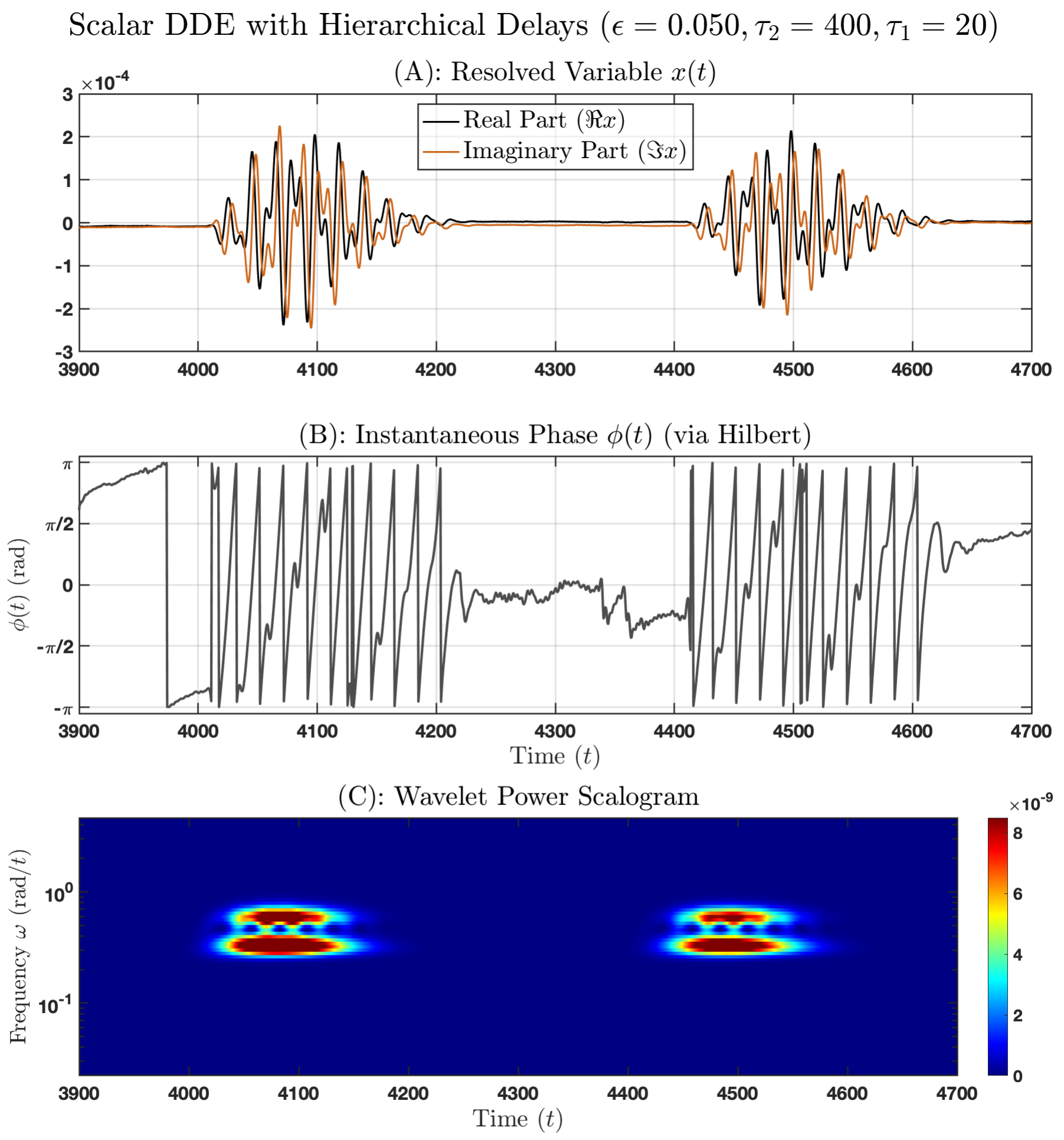}
\caption{\textbf{Emergent AM-FM Dynamics in the Scalar Hierarchical DDE.} Here $a=-0.4 + 0.5i$, $b=0.1$, $c=0.3$, $\tau_1=20$ and $\tau_2=400$ ($\epsilon=5\times 10^{-2}$) for the deterministic scalar model (Eq.~\eqref{Eq_HDM}). Panels A and B display the real/imaginary trajectories (see legend) and the highly non-periodic instantaneous phase obtained via Hilbert transform \cite{huang2014hilbert}, respectively. Panel C introduces the wavelet power scalogram, which serves as the primary diagnostic tool for these transient dynamics. The scalogram visually captures the AM-FM signature as distinct, localized ``islands'' situated near a fixed frequency band. The punctuated bursting and fading of these structures in time reflects Amplitude Modulation, while their complex, broadened shape in the time-frequency domain manifests the active Frequency Modulation. The gentle, long-term vertical drift observed is a secular shift driven strictly by the weak, differential damping of the interacting modes.}
\label{Fig_DDEtoy_Phase_Scalogram}
\vspace{-2ex} 
\end{figure}

The continuous interaction of \textit{many} such pairs, each with slightly different decay rates $\sigma_k$ and frequencies $\omega_k$, results in a  complex, non-repeating pattern of constructive and destructive interference: an \textit{emergent beat pattern}. The dense spectral hierarchy $\Sigma$, generated strictly by the disparate delays, is thus the mathematical engine responsible for stable, long-lived, and non-repeating AM-FM dynamics.

To visualize and quantify these emergent AM-FM dynamics, we employ the continuous wavelet transform (CWT) scalogram. Figure \ref{Fig_DDEtoy_Phase_Scalogram} presents a simulation of the scalar model (Eq.~\eqref{Eq_HDM}) set within a high spectral density regime. The resulting wavelet scalogram (Fig.~\ref{Fig_DDEtoy_Phase_Scalogram}C) provides a stunning time-frequency portrait of the $M$-mode interactions. Rather than a continuous, monochromatic frequency band, the energy manifests as distinct, localized ``islands'' situated near a fixed frequency. 

The punctuated nature of these features—bursting and fading in time—is the direct signature of constructive and destructive AM. Simultaneously, the mere existence of these complex island structures, rather than pure tonal pulses, is the manifestation of the active FM. Driven by the phase cross-terms of the dense, near-degenerate modes, the FM broadens and shapes the localized energy within the time-frequency domain. Furthermore, it is important to note that these wave packets do not arbitrarily wander across the frequency axis; any long-term, gentle drift of the dominant frequency band is a secular shift dictated strictly by the weak, differential damping ($\delta \sigma_k$) of the interacting modes.

Because the wavelet scalogram so perfectly isolates these transient beat patterns, it will serve as the central diagnostic tool for the remainder of this paper. We will leverage this time-frequency representation to design and validate data-driven inverse stochastic models, first applying it to recover the dynamics of this theoretical system, and ultimately utilizing it to uncover the latent, highly-coupled AM-FM beat patterns governing the multi-scale GOES-16 cloud field observations.



\subsection*{Inverse Stochastic Modeling and Wavelet-Based Penalization}

\noindent Transitioning from the theoretical generation of AM-FM dynamics to data-driven inverse modeling presents a profound challenge. For systems governed by sensitive, dense spectral geometries---such as the hierarchical delay models developed above---traditional regression techniques based on point-by-point trajectory matching (e.g., Mean Squared Error) inevitably fail. In AM-FM regimes, a slight parameter error or stochastic perturbation may induce a phase shift in the beat pattern \cite{hoppensteadt1998thalamo}. When a simulated burst is evaluated point-by-point against a target burst that is even slightly out of phase, the $L_2$ error diverges catastrophically, blinding the optimization algorithm to models that are dynamically and physically accurate.

To overcome this issue, we adopt a structural and spectral matching philosophy centered on wavelet-based penalization. Rather than forcing the model to reproduce the exact chronological realization of a chaotic target, the objective is to learn a stochastic system whose generated trajectories belong to the exact same statistical and spectral manifold as the observations, as quantified by their time-frequency wavelet signatures. To demonstrate this methodology before scaling to the high-dimensional GOES-16 cloud observations, we deploy a Bayesian optimization framework to invert the theoretical scalar DDE testbed:
\be\label{Eq_HDM_Stoch}
\mathrm{d}x(t) = \big(a x(t) + b x(t-\tau_1) + c x(t-\tau_2)\big)\mathrm{d}t + \sigma \mathrm{d}W_t,
\ee
where the hierarchical delays are configured in the high spectral density regime ($\epsilon=0.05$), yielding true underlying delays of $\tau_1 = 20$ and $\tau_2 = 400$, with true linear coefficients $a=-0.4+0.5i$, $b=0.1$, and $c=0.3$. 

To illustrate our learning approach, we assume the instantaneous baseline dynamics is known and fix $a=-0.4+0.5i$ in Eq.~\eqref{Eq_HDM_Stoch}. Thus, the inverse problem requires learning the continuous delayed feedback coefficients ($b, c$), the precise hierarchical delay structure ($\tau_1, \tau_2$), and the magnitude of the stochastic noise ($\sigma$) strictly from a highly limited observational window. In this testbed, the model is trained exclusively on a single transient AM-FM burst spanning $t $ in $ [3900, 4300]$.

The optimization landscape of delay differential equations is famously non-convex, characterized by numerous local minima due to the periodic nature of delayed feedback \cite{sandoz2023sindy}. To navigate this topology, we employ Bayesian optimization utilizing an Expected Improvement acquisition function \cite{snoek2012practical,gelbart2014bayesian}. By modeling the objective function with a Gaussian Process (GP) surrogate \cite{rasmussen2006gaussian}, the algorithm allows us to jointly search the parameter space of the delays $\tau_{1,2}$ (bounded linearly) and the stochastic noise magnitude $\sigma$ (bounded logarithmically).

To ensure computational tractability, the optimization relies on a nested architecture. For every candidate pair of delays $(\tau_1, \tau_2)$ proposed by the Bayesian outer loop, the inner loop computes the optimal instantaneous linear coefficients ($b, c$) via regularized OLS on the gradient of the target training window. Once the deterministic skeleton is defined, the framework simulates an extended, frozen-seed realization of the candidate Stochastic DDE via a high-resolution Euler-Maruyama scheme.

The candidate trajectory is then evaluated using a composite structural cost function, $J$, computed in the time-frequency domain via continuous wavelet transforms (CWTs). Time-series approaches based on the wavelet transform and analytic signals  have shown their ability in characterizing non-stationary behavior and elucidating phase interactions \cite{ivanov1996scaling}. This makes CWTs ideally suited to capture the highly coupled, multi-frequency phase dynamics of our system, such as those encapsulated by the instantaneous phase derivative; see e.g.~Eq.~\eqref{Eq_dot_phi}. 

The cost function balances three primary statistical moments:
\be
J = \beta J_{\text{var}} + (1-\beta) \left( \frac{1}{2} J_{\text{gws}} + \frac{1}{2} J_{\text{kurt}} \right),
\ee
where $\beta = 0.80$ aggressively prioritizes the recovery of the total signal variance ($J_{\text{var}}$). The spectral architecture is penalized by $J_{\text{gws}}$, computed as the mean squared error in the logarithmic domain of the Global Wavelet Spectrum (GWS), ensuring broadband frequency matching across orders of magnitude. Finally, $J_{\text{kurt}}$ measures the logarithmic error of the spectral kurtosis, explicitly forcing the optimizer to match the intermittency and ``burstiness'' characteristic of AM-FM temporal localization.

\begin{figure}
    \centering
    \includegraphics[width=1\linewidth, height=0.5\textwidth]{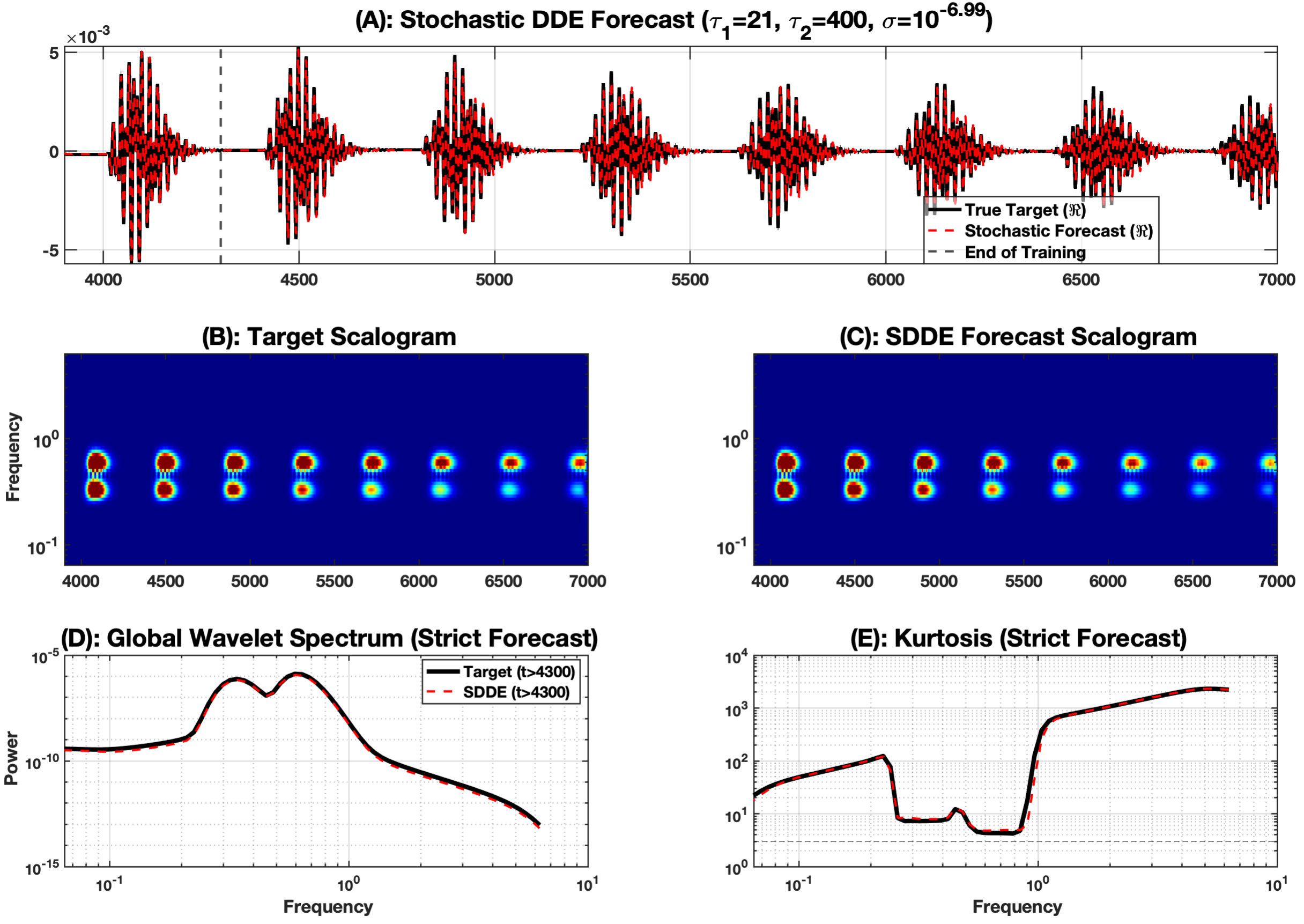}
    \caption{\textbf{Inverse Stochastic Modeling of the Scalar Hierarchical DDE.} The model was trained exclusively on a single energetic burst within the training window ($t < 4300$), optimizing the structural cost function via Bayesian inference. \textbf{(A)}: The resulting out-of-sample stochastic forecast (red) successfully generates the long-term, non-periodic AM-FM beat patterns of the true deterministic target (black). \textbf{(B-C)}: The wavelet power scalograms demonstrate excellent morphological agreement, confirming that the learned model ($\tau_1=21$, $\tau_2=400$, $\sigma=10^{-6.99}$) captures the transient, highly localized energetic ''islands'' characteristic of the underlying spectral hierarchy. \textbf{(D-E)}: Strict out-of-sample evaluation ($t > 4300$) of the Global Wavelet Spectrum and Spectral Kurtosis proves the model perfectly reconstructs the broadband frequency distribution and the non-Gaussian intermittency of the target signal.}
    \label{Fig_Inverse_Toy}
    \vspace{-0.1cm}
\end{figure}

We analyze now the numerical results. 
The Bayesian optimization converged successfully within 60 objective evaluations, perfectly identifying the underlying high-density parameters from the single training burst. The algorithm recovered almost exactly the hierarchical delays ($\tau_1 = 21, \tau_2 = 400$) and accurately estimated the deterministic coefficients ($b=0.10, c = 0.29$) alongside a near-zero stochastic noise floor ($\sigma \approx 10^{-6.99}$), correctly inferring the fundamentally deterministic origin of the target testbed.

Figure~\ref{Fig_Inverse_Toy} details the generative performance of the inverted model. Despite being trained on only a fraction of the system's dynamical phase space ($3900<t < 4300$), the stochastic emulator gracefully transitions into an extended out-of-sample forecast (Fig.~\ref{Fig_Inverse_Toy}A) that successfully reproduces the non-periodic, macroscopic beat patterns spanning the few hundreds of time units. 

Crucially, the scalogram of the forecast (Fig.~\ref{Fig_Inverse_Toy}C) remains structurally indistinguishable from the target scalogram (Fig.~\ref{Fig_Inverse_Toy}B). The inverse model actively generates the transient, localized energetic bursts associated with the coupled amplitude and frequency modulation derived analytically in the preceding section. This visual confirmation is supported by the out-of-sample empirical metrics (Fig.~\ref{Fig_Inverse_Toy}D,E), which show an almost flawless overlap in the Global Wavelet Spectrum and the Spectral Kurtosis.

By proving that this spectral-matching Bayesian philosophy can successfully invert highly sensitive, dense-spectrum AM-FM dynamics in a controlled scalar setting, we establish the robust methodological foundation necessary to deploy this exact framework onto the high-dimensional, empirically derived multi-scale matrices of the GOES-16 cloud observations.


\begin{figure*}
\centering
\includegraphics[width=0.98\linewidth, height=.55\textwidth, keepaspectratio]{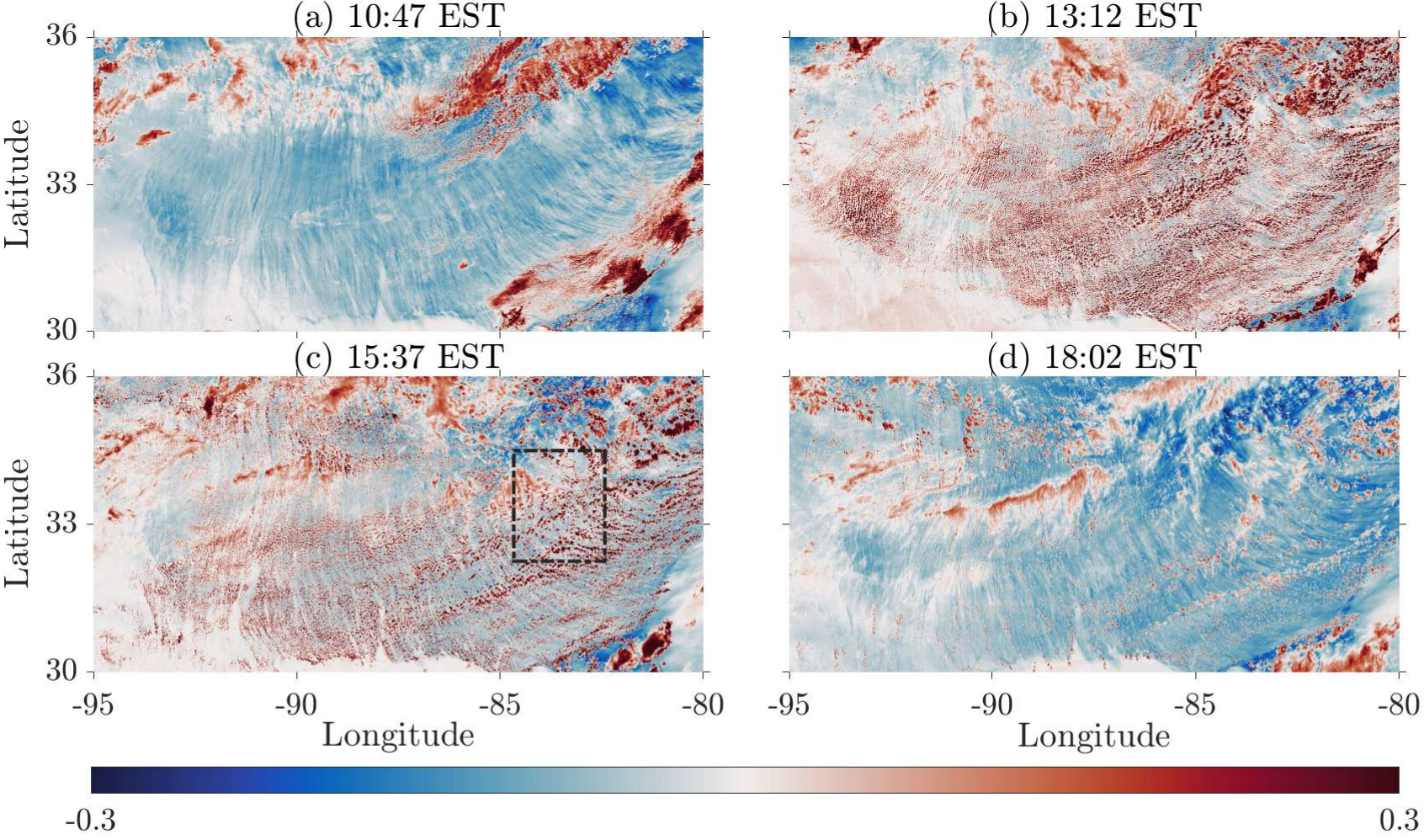}
\caption{{\bf High-resolution (corrected) reflectance anomaly field of the diurnal cloud evolution}. These snapshots are obtained from the GOES-16 ABI's ``Red'' visible band (channel 2; $0.64\,\mu\mathrm{m}$); see \textit{Methods} section. The dashed box highlights a region where individual clouds continuously form and dissipate like ``pearls on a string'' along the broader mesoscale cloud streets that form parallel to the advection wind \cite{Dror2021}. Many such regions exist throughout the domain. The snapshots represent distinct stages of the semidiurnal cycle: an early formation stage at 10:47 EST (a), peak cloud field deck organization at 13:12 (b) and 15:37 EST (c), and a dissipation event at 18:02 EST (d).}
\label{Fig_intro}
\end{figure*}


\subsection*{The Diurnal Cloud Field Organization Challenge}
To demonstrate the framework's capacity to resolve the spectral structures underlying beat patterns in real-world systems, we turn to a notoriously challenging problem: the characterization and simulation of organized patterns formed by shallow convective clouds \cite{nair_1998, heus_2013, stevens_2019, dror_2020}. Low clouds play a critical role in Earth's radiation budget and overall climate sensitivity \cite{bony_2005, webb_2006, vial_2017, nuijens_2019}, yet their transient mesoscale organization is poorly represented in global models. We focus on a specific subset known as "Green-Cumulus" (GreenCu), characterized by distinct street or grid-like spatial patterns \cite{dror_2020, Dror2021}. 

Figure~\ref{Fig_intro} displays a segment in the life of the target cloud field, captured via high-resolution satellite data from the GOES-16 Advanced Baseline Imager (channel 2, $0.64\,\mu m$) \cite{schmit_2017}. Our model's foundation rests strictly on this limited daytime dataset (8 hours of evolution comprising 97 snapshots at 5-minute intervals). This strict restriction is physically crucial: tracking these highly organized cloud patterns through the night would necessitate shifting to infrared data, introducing severe complexities from mixing fundamentally different radiance origins (infrared vs. visible range).

\begin{figure}[ht!]
\centering
    \includegraphics[width=1\linewidth, height=0.35\textwidth]{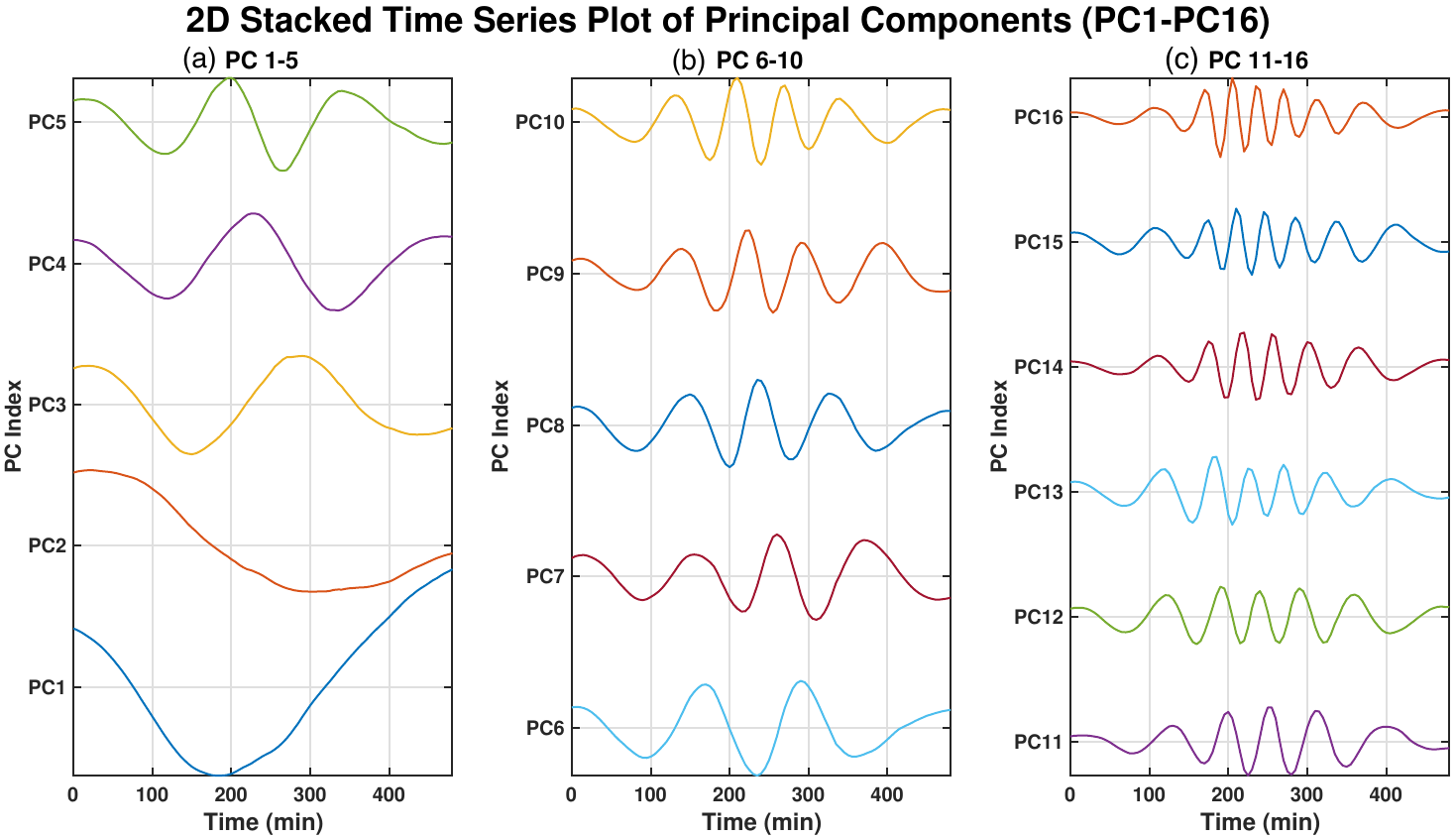}
\caption{\textbf{2D Stacked Time Series Plot of Principal Components.} The first 16 PCs obtained from EOF decomposition following  \cite{Dror2021}, of the GOES-16 GreenCu dataset; see \textit{Methods} section. Note the AM-FM behavior, specially pronounced for the PC5 to PC16.}
\label{Fig_PCs}
\vspace{-3.5ex}
\end{figure}

Our modeling takes place in the Empirical Orthogonal Function (EOF) space \cite{lorenz1956empirical,preisendorfer1988principal} of this reflectance anomaly field, following the methodology of \cite{Dror2021}. The resulting Principal Components (PCs) serve as our derived, low-dimensional \textit{latent variables} for capturing the essential cloud dynamics. For this dataset, we retain the $d=16$ leading PCs, capturing approximately $60\%$ of the variance. As documented in previous spatial analyses, this specific truncation is physically motivated: it provides a sufficient hierarchical basis to encompass the full cascade of relevant dynamics, from the macroscopic boundary conditions down to the coherent formation and dissipation of individual clouds \cite{Dror2021}. 

Figure~\ref{Fig_PCs} illustrates the temporal evolution of these 16 PCs. The empirical challenge is immediately apparent: the time series are fundamentally oscillatory but exhibit pronounced AM-FM behavior. Specifically in the deeper modes ($\text{PC}_{5}$ through $\text{PC}_{16}$), the signals display continuous, state-dependent frequency shifts and intermittent bursts of localized energy. 

Given that this phenomenon is transient, short-lived, and strictly tied to the diurnal cycle, traditional long-term prediction of independent events is ill-posed. Instead, the MSHDM functions as a high-fidelity stochastic emulator designed for a ``Groundhog Day'' scenario: it generates massive synthetic ensembles of the observed event, where each realization captures a new stochastic instance of that day's dynamics. By producing coherent spatio-temporal dynamics consistent with the limited record, this emulator enables us to glean fundamental dynamical insights—particularly the latent AM-FM beat patterns—that are otherwise statistically inaccessible from the raw dataset.

Scaling the hierarchical delay methodology from the scalar test case to this higher-dimensional empirical system, however, introduces its own curse of dimensionality. Capturing the dense cross-scale interactions between 16 PCs---each paired with respective delayed memories---would produce a massive, highly parameterized regression matrix. To resolve this while preserving the capacity to generate the observed AM-FM beat patterns, the multivariate model explicitly embeds dimensionality reduction directly into its dynamical structure.

\subsection*{The Tripartite MSHDM for Cloud Field Organization}
The fundamental advantage of the proposed Multilayered Stochastic Hierarchical Delay Model (MSHDM) lies in its structural simplicity and broad applicability to organized, multiscale dynamical fields, such as cloud reflectance. This is achieved through a synergistic combination of topological subsystem splitting and Partial Least Squares (PLS) dimensionality reduction. Following the general layered architecture defined in Eq.~(\ref{Eq_MSHDM_General}), let $\mathbf{u}(t) \in \mathbb{R}^d$ denote the state vector of PCs extracted from the observed field. Guided by the characteristic timescales captured by this underlying system of PCs \cite{Dror2021}, we decompose the state vector into three topologically distinct but coupled subsystems ($L=3$):
\begin{equation}
    \mathbf{u}(t) = \begin{bmatrix} \mathbf{s}(t) & \mathbf{x}(t) & \mathbf{y}(t) \end{bmatrix}^T,
\end{equation}
where $\mathbf{s}(t) \in \mathbb{R}^m$ represents the dominant, large-scale cyclical envelope; $\mathbf{x}(t) \in \mathbb{R}^p$ encapsulates the intermediate, slow mesoscale dynamics; and $\mathbf{y}(t) \in \mathbb{R}^q$ captures the fast, high-frequency turbulent modes, such that $m + p + q = d$, the total number of retained dimensions.

A direct estimation of the dense cross-layer and discrete delay matrices required by the general formulation (Eq.~\eqref{Eq_MSHDM_General}) via Ordinary Least Squares would be mathematically ill-posed given the severely limited duration of the GOES-16 cloud dataset, inevitably resulting in catastrophic overfitting. To ensure a stable, generalizable solution that accurately reflects the causal flow of energy across these scales, it is essential to employ robust regularization methods that restrict the learning process to a limited number of effective degrees of freedom \cite{hastie2009linear}. Rather than simply penalizing the matrix norm (like Ridge or LASSO regression), our framework utilizes PLS \cite{abdi2003partial}. PLS implicitly bottlenecks the high-dimensional concatenated history and cross-layer vectors through an optimal, highly compact latent subspace of rank $r$ before mapping them to the local state tendency. This structural dimensionality reduction is encapsulated by the projection operator $\mathbf{\Phi}$, yielding the explicitly coupled, tripartite Stochastic Differential Delay Equation (SDDE):
\begin{widetext}
\begin{align}
    \mathrm{d}\mathbf{s}(t) &= \Big[ \mathbf{A}_{\mathbf{s}} \mathbf{s}(t) + F\Big] \mathrm{d}t + \mathbf{\Sigma}_{\mathbf{s}} \mathrm{d}\mathbf{W}_{\mathbf{s}}(t), \label{eq:sdde_solar} \\
    \mathrm{d}\mathbf{x}(t) &= \Big[ \mathbf{A}_{\mathbf{x}} \mathbf{x}(t) + \mathbf{\Phi}_{\mathbf{x}} \begin{bmatrix} 1 & \mathbf{y}(t) & \mathbf{x}(t-\tau_1^{(1)}) & \mathbf{x}(t-\tau_2^{(1)}) & \mathbf{s}(t) \end{bmatrix}^T \Big] \mathrm{d}t + \mathbf{\Sigma}_{\mathbf{x}} \mathrm{d}\mathbf{W}_{\mathbf{x}}(t), \label{eq:sdde_slow} \\
    \mathrm{d}\mathbf{y}(t) &= \Big[ \mathbf{A}_{\mathbf{y}} \mathbf{y}(t) + \mathbf{\Phi}_{\mathbf{y}} \begin{bmatrix} 1 & \mathbf{x}(t)  & \mathbf{y}(t-\tau_1^{(2)}) & \mathbf{y}(t-\tau_2^{(2)}) & \mathbf{s}(t) \end{bmatrix}^T \Big] \mathrm{d}t + \mathbf{\Sigma}_{\mathbf{y}} \mathrm{d}\mathbf{W}_{\mathbf{y}}(t). \label{eq:sdde_fast}
\end{align}
\end{widetext}

Here, the parameter pairs $(\tau_1^{(1)}, \tau_2^{(1)})$ and $(\tau_1^{(2)}, \tau_2^{(2)})$---which we hereafter denote more intuitively as $(\tau_{\mathrm{slw}_1}, \tau_{\mathrm{slw}_2})$ and $(\tau_{\mathrm{fst}_1}, \tau_{\mathrm{fst}_2})$---dictate the specific historical memory bounds for the slow and fast blocks, respectively.
The matrices $\mathbf{A}_{\mathbf{s}}$, $\mathbf{A}_{\mathbf{x}}$, and $\mathbf{A}_{\mathbf{y}}$ govern the instantaneous (lag-0) intra-layer Markovian interactions (corresponding to $\mathbf{A}_0^{\ell\ell}$ in the general notation of Eq.~\eqref{Eq_MSHDM_General}), while the PLS operators $\mathbf{\Phi}_{\mathbf{x}}$ and $\mathbf{\Phi}_{\mathbf{y}}$ manage all cross-layer and delayed couplings. Notably, the inclusion of the constant scalar $1$ at the head of each predictor vector allows $\mathbf{\Phi}$ to elegantly absorb the geometric intercept, shifting the underlying stochastic attractor \cite{csg11} to naturally capture any background asymmetric drift or non-zero means in the observations. 

Finally, $\mathrm{d}\mathbf{W}(t)$ represents the differential increments of mutually independent multidimensional Wiener processes (i.e., Gaussian white noise), and $\mathbf{\Sigma}$ represents the lower Cholesky factors of the stochastic innovation covariances. To reflect the topological separation, the global noise covariance is block-diagonal; however, within each specific subsystem, the covariance matrix is fully dense, explicitly preserving the instantaneous cross-modal correlations among the constituent PCs. 


\subsection*{Hierarchical Forcing and Markovian Constraints}

This explicit tripartite topology mathematically enforces the separation of distinct physical processes through targeted constraints on the lag-0 matrices.

The first subsystem is that of the cyclical envelope ($\mathbf{s}$).
The $\mathbf{s}$-equation (Eq.~\eqref{eq:sdde_solar}) models the primary forcing as an autonomous system. To properly capture resonant harmonic frequencies, the dimension $m$ is constrained to an even integer. The Markovian lag-0 matrix $\mathbf{A}_{\mathbf{s}}$ is not learned freely; rather, it is strictly constrained during optimization to a block-diagonal operator composed of $2 \times 2$ antisymmetric submatrices, representing slightly dissipative oscillator pairs:
\begin{equation}
    \mathbf{A}_{\mathbf{s}}^{(k)} = \begin{bmatrix} \gamma_k & \omega_k \\ -\omega_k & \gamma_k \end{bmatrix}, \quad \gamma_k < 0,
\end{equation}
where $\omega_k$ dictates the primary angular frequency of the $k$-th mode pair. 

The second and third subsystems (Eqns.~\eqref{eq:sdde_slow}-\eqref{eq:sdde_fast}) correspond to the driven slow and fast manifolds ($\mathbf{x}$ and $\mathbf{y}$). 
Unlike the cyclical envelope, the instantaneous lag-0 matrices for the slow and fast subsystems ($\mathbf{A}_{\mathbf{x}}$ and $\mathbf{A}_{\mathbf{y}}$) are learned as unconstrained, dense operators, allowing for complex, non-rotational energy transfers between the coupled meso-scale and turbulent modes. Furthermore, while guided by their own localized delay memories, their augmented PLS predictors $\mathbf{\Phi}_{\mathbf{x}}$ and $\mathbf{\Phi}_{\mathbf{y}}$ explicitly ingest the contemporaneous state of the cyclical envelope, $\mathbf{s}(t)$. This mechanism physically wires the large-scale thermodynamic boundaries to instantaneously dictate the behavior of the subordinate hierarchical subsystems.

\medskip
\subsection*{Hybrid Ergodic Hyperparameter Optimization}
To prevent overfitting the model's structural elements to the limited observational record, the model hyperparameters---namely the delay terms and the PLS latent subspace ranks ($r$) for each block---are calibrated using a hybrid, topology-aware objective function. The optimization strictly separates deterministic trajectory tracking from stochastic ergodic consistency. 

During calibration, the deterministic linear components ($\mathbf{A}$) are fit first, followed by the PLS regression ($\mathbf{\Phi}$) on the residuals. The empirical covariance of the final PLS residuals is then decomposed via Cholesky factorization to populate the noise matrices ($\mathbf{\Sigma}$). This provides an empirical data-driven closure, ensuring that the stochastic innovations are explicitly calibrated to the variance of the residual modeling error for each subsystem, thereby introducing realistic ensemble variability into the simulations.

To evaluate the topological costs, the full SDDE is simulated over both the in-sample training window and an extended out-of-sample forecasting horizon. The objective function explicitly separates deterministic trajectory tracking from stochastic ergodic consistency. 

To quantify the deterministic trajectory tracking, we define an in-sample cost, $J_{\mathbf{s}}$. The large-scale envelope $\mathbf{s}(t)$ functions as the system's underlying clock; consequently, $J_{\mathbf{s}}$ evaluates strictly in-sample performance, penalizing phase and amplitude drift via a variance-normalized Mean Squared Error over the training horizon $N_{\text{train}}$:
\begin{equation}
    J_{\mathbf{s}} = \frac{1}{m N_{\text{train}}} \sum_{j=1}^{m} \sum_{t=1}^{N_{\text{train}}} \frac{\big( s_{j}^{\text{sim}}(t) - s_{j}^{\text{obs}}(t) \big)^2}{\text{Var}(s_{j}^{\text{obs}}) + \epsilon},
\end{equation}
where $j$ indexes the dimensions of the cyclical subsystem and $\epsilon$ is a small regularization constant to ensure numerical stability.

To enforce the stochastic ergodic consistency, we introduce the out-of-sample costs, $J_{\mathbf{x}}$ and $J_{\mathbf{y}}$, for the remaining layers. Because the slow and fast subsystems are dominated by chaotic, turbulent physics where pointwise trajectory matching is mathematically ill-posed, their performance is evaluated strictly \textit{out-of-sample} over the extended integration horizon. Empirical observations of the principal components indicate that the temporal dynamics naturally organize into coupled oscillatory pairs exhibiting near phase-quadrature behavior (e.g., PC3 and PC4, PC5 and PC6). To leverage this physical structure and ensure computational efficiency during calibration, the ergodic penalties for a given subsystem $\gamma$ in $\{\mathbf{x}, \mathbf{y}\}$ are evaluated pairwise.

Unlike the calibration of the theoretical toy model, we deliberately exclude the spectral kurtosis penalty for the cloud application. As a fourth-order statistical moment, kurtosis is notoriously sensitive to finite sample sizes. Estimating it from our strictly limited observational window (comprising merely 97 discrete data points per channel) would be statistically unreliable and introduce spurious noise into the optimization rather than meaningful structural guidance.

Instead, for a subsystem consisting of $K_\gamma = d_\gamma/2$ coupled pairs, the overall subsystem cost is formulated as the average convex combination of the variance and GWS errors across its pairs:
\begin{equation}
    J_{\gamma} = \frac{1}{K_\gamma} \sum_{k=1}^{K_\gamma} \Big[ \beta J_{\text{var}, k} + (1 - \beta) J_{\text{gws}, k} \Big],
\end{equation}
where $\beta$ in  $[0, 1]$ provides a strict physical trade-off between capturing the absolute energy magnitude and preserving the temporal-frequency content. 

The variance cost $J_{\text{var}, k}$ penalizes macroscopic amplitude collapse or artificial blow-up in the out-of-sample simulation for the $k$-th pair. By exploiting the phase-quadrature symmetry, this cost is computed robustly using the primary constituent component of the pair, denoted $u_k^*$:
\begin{equation}
    J_{\text{var}, k} = \left( \frac{\text{Var}(u_{k}^{*,\text{sim}}) - \text{Var}(u_{k}^{*,\text{obs}})}{\text{Var}(u_{k}^{*,\text{obs}}) + \epsilon} \right)^2.
\end{equation}

Similarly, the spectral cost $J_{\text{gws}, k}$ ensures the correct distribution of energy across timescales for the pair. Let $W_k^*(\omega)$ denote the GWS power of the representative component at frequency $\omega$, evaluated over a discrete target frequency grid $\Omega$. To heavily penalize order-of-magnitude distortions in the spectral cascade, the loss is computed as the Mean Squared Error of the logarithmic spectra:
\begin{equation}
    J_{\text{gws}, k} = \frac{1}{|\Omega|} \sum_{\omega \in \Omega} \left[ \log_{10}\left( \frac{\widetilde{W}_{k}^{*,\text{sim}}(\omega) + \epsilon}{W_{k}^{*,\text{obs}}(\omega) + \epsilon} \right) \right]^2,
\end{equation}
where $\widetilde{W}_{k}^{*,\text{sim}}(\omega)$ is the simulated out-of-sample GWS interpolated onto the target observational frequency grid $\Omega$.

The can then design the total objective function. 
To ensure that deep memory structures are only retained if they demonstrably improve the spectral physics without degrading the primary deterministic forcing, the final optimization objective integrates the three topological costs with equal weight. To simultaneously guarantee numerical stability and effective optimization across the parameter space, this total objective is formulated using a logarithmic transformation with a regularization constant $\epsilon_2 = 10^{-6}$:
\begin{equation}
    J_{\text{total}} = \log_{10}\left( \frac{1}{3} J_{\mathbf{s}} + \frac{1}{3} J_{\mathbf{x}} + \frac{1}{3} J_{\mathbf{y}} + \epsilon_2 \right).
\end{equation}
 
  Crucially, calibrating a tripartite stochastic delay model across a 6-dimensional mixed hyperparameter space---combining the four physical delay bounds ($\tau_{\mathrm{slw}_1}, \tau_{\mathrm{slw}_2}, \tau_{\mathrm{fst}_1}, \tau_{\mathrm{fst}_2}$) with the integer PLS latent dimensions for the slow and fast subsystems---generates a highly complex and rough error landscape. Suboptimal combinations of delays or insufficient latent complexities can yield poorly conditioned regression operators ($\mathbf{\Phi}$) or a DDE spectrum with unstable modes, leading to catastrophic trajectory divergence during the out-of-sample numerical integration. Optimizing a raw linear error in this context would cause the Gaussian Process (GP) surrogate model to be entirely dominated by the steep gradients of these failure regimes. The logarithmic transformation gracefully compresses these potentially massive variance scales, enabling the GP kernel to effectively resolve and differentiate the subtle topographies of the highly performing, low-error regions, while $\epsilon_2$ strictly bounds the argument.

Recent advancements in data-driven delay modeling have similarly recognized the computational advantages of Bayesian optimization over brute-force grid searches to navigate complex parameter spaces. For instance, extensions of the SINDy framework successfully deploy Bayesian algorithms to identify optimal discrete lag times \cite{sandoz2023sindy, pecile2025data}. A fundamental distinction, however, lies in the optimization objective. These frameworks typically identify delays by minimizing an in-sample, pointwise derivative reconstruction error. 
While this is a highly proven and effective philosophy for discovering explicit governing equations, calibrating stochastic delay models from severely short observational records introduces distinct optimization priorities. In such limited-data regimes, intrinsic stochastic fluctuations naturally generate large instantaneous residuals. Rather than relying on pointwise trajectory matching---which necessitates careful regularization to navigate this noise---an alternative approach is to optimize directly for the system's macroscopic statistical features. By fundamentally decoupling the objective---evaluating the solar deterministic clock in-sample while strictly enforcing out-of-sample spectral and ergodic consistency (via the GWS) for the other stochastic subsystems—our framework leverages Bayesian optimization  to explicitly target the underlying AM-FM structural geometry. Relying on these spectral metrics prioritizes long-term structural stability, providing a complementary approach designed specifically to recover emergent wave properties from short data bursts.

To serve our purposes, the hyperparameter search was executed using Bayesian optimization driven by an Expected Improvement (EI) acquisition function. The EI metric intelligently navigates this 6-dimensional space by balancing two competing goals:
\begin{itemize}
    \item Exploitation: Sampling where the GP surrogate predicts a very low cost, thereby refining the search at the absolute bottom of a known performance valley.
    \item Exploration: Sampling regions where the GP is highly uncertain, probing unmapped areas of the parameter space to ensure a deeper, globally optimal basin is not missed.
\end{itemize}

While the Bayesian optimization effectively resolves the complex temporal structure of the SDDE (the optimal delays and latent PLS dimensions), guaranteeing out-of-sample robustness requires an additional tuning step. Specifically, the raw empirical covariance of the PLS residuals—used to initially populate the noise matrices ($\mathbf{\Sigma}$)—is a purely in-sample estimate. During an extended out-of-sample integration, imbalances between the learned deterministic damping and these raw stochastic innovations can compound, leading to macroscopic energy drift. To enforce strict long-term energy conservation, the optimal model undergoes a final post-optimization tuning of its noise intensities, rooted in the fluctuation-dissipation theorem; see \textit{Methods: Iterative Calibration of Stochastic Forcing}.

\medskip
\subsection*{Intrinsic Timescales and $L_1$-Norm Decomposition of the Slow-Fast Drivers}

To understand the dynamic underpinnings of the cloud field across the slow and fast components, we first establish the intrinsic baseline oscillations of the system before examining the PLS regression weights mapping the high-dimensional predictor spaces to the right-hand sides of both the slow and fast equations.


When interpreting these physical drivers, it is critical to map the temporal PCs to their corresponding spatial EOFs, which represent progressively smaller spatial scales \cite{Dror2021}. In the context of shallow cumulus convection, our $d=16$ truncation allows us to partition the dynamics into three distinct physical regimes. The leading low-frequency modes ($\text{PC}_{1}$--$\text{PC}_{2}$) capture the macroscopic, slow-moving boundary conditions of the cloud field. The intermediate-frequency modes ($\text{PC}_{3}$--$\text{PC}_{10}$) resolve mesoscale organization. In previous spatial analyses of this dataset \cite{Dror2021}, this intermediate band was shown to harbor a near-periodic $1.5$~h oscillation (manifested in $\text{PC}_{7}$--$\text{PC}_{10}$ as explained below), interpreted as the fingerprint of convective IGWs traveling throughout the domain. Finally, the high-frequency modes ($\text{PC}_{11}$--$\text{PC}_{16}$) are mathematically required to capture the rapid, localized formation and dissipation of individual clouds lined up in strips. This highly organized grid-like pattern is typically observed in cumulus convection over land \cite{plank1966wind, plank1969size, da2011cloud, Dror2021}.

Before evaluating the cross-scale structural weights, it is important to acknowledge the intrinsic dynamics of the macroscopic forcing ($\text{PC}_{1}$--$\text{PC}_{2}$). Because the solar subsystem is estimated independently (Eq.~\ref{eq:sdde_solar}), its instantaneous linear matrix ($\mathbf{A}_{s}$) autonomously extracts a period of $11.82$~hours from its eigenvalues. The fundamental semidiurnal daylight harmonic is thus natively captured. Conversely, for the adjacent macroscopic modes of the slow group (Pairs 2 and 3), the 8-hour limit of the observational data creates a fundamental resolution barrier. Because power spectrum estimation  requires multiple full cycles to confidently identify a spectral peak, it is statistically impossible to resolve their native dominant periods directly from the empirical observations without severe edge-effect distortion. We show in the next section our MSHDM modeling enables us to overcome this observational ``no man's land.''

To decipher the high-dimensional forcing that interacts with these baseline oscillations, we analyze the PLS weights across this specific structural hierarchy. By performing an $L_1$-norm decomposition---aggregating the absolute weights ($\sum |w|$) of the PLS coefficients into distinct physical categories for each scale---we can trace how energy and structural information cascade between the macroscopic boundaries, the mesoscale gravity waves, and the individual cloud cells. Unpacking these total deterministic weights into stacked volume representations brings the underlying physics governing the energy exchange and internal memory of the cloud field model directly into view.

\begin{figure}[htbp]
    \centering
     \includegraphics[width=1\linewidth, height=0.5\textwidth]{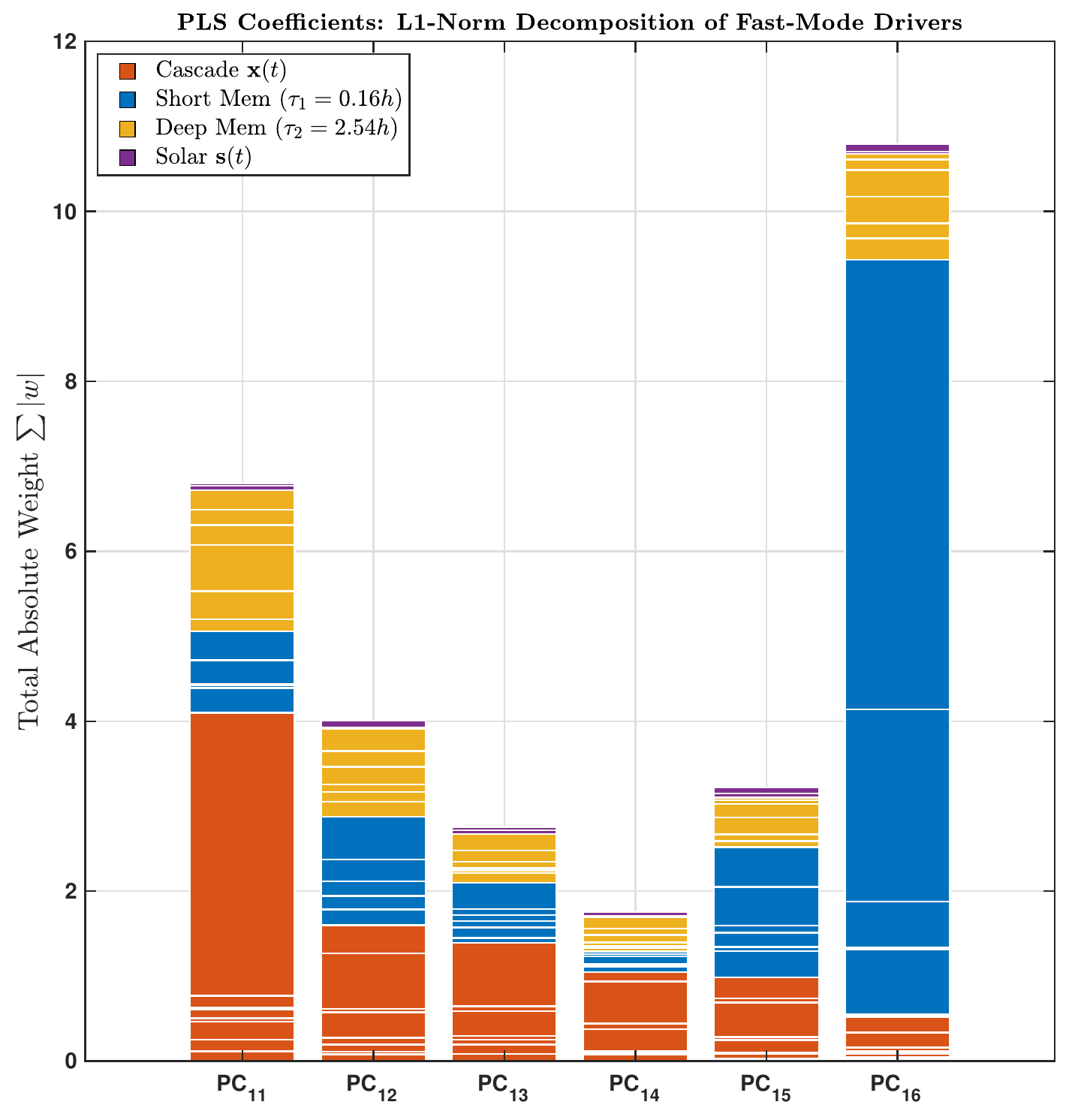}
    \caption{\textbf{$L_1$-Norm Decomposition of Fast-Mode Predictors.} The stacked bar chart illustrates the total absolute deterministic weight ($\sum |w|$) applied to the right-hand side of each fast-mode equation ($\text{PC}_{11}$--$\text{PC}_{16}$). The predictors are aggregated into four distinct physical mechanisms: the instantaneous forward energy cascade from the intermediate slow modes (orange), short-term convective mixing memory at $\tau_{\mathrm{fst}_1} = 0.16$~h (blue), deeper environmental memory at $\tau_{\mathrm{fst}_2} = 2.54$~h (yellow), and instantaneous solar forcing (purple). The distinct U-shaped curve highlights a regime where energy injection ($\text{PC}_{11}$) and rapid mixing dissipation ($\text{PC}_{16}$) are heavily deterministically forced, while the intermediate individual cloud modes are dominated by stochastic variability.}
    \label{PLS_L1-norm_fast}
\end{figure}

With this structural context established, the $L_1$-norm decomposition of the fast-mode drivers (Fig.~\ref{PLS_L1-norm_fast}) yields several critical insights into how the model mathematically manages localized convective lifecycles. First, the boundary cascade is visually confirmed at the leading edge of the fast spectrum. The total deterministic volume of $\text{PC}_{11}$ is heavily dominated by the cross-coupled instantaneous slow modes ($\mathbf{x}(t)$). Sitting directly adjacent to the prescribed spectral boundary, $\text{PC}_{11}$ is primarily driven by energy bleeding across from the mesoscale organization---most notably from $\text{PC}_{10}$, the fastest of the intermediate modes, which exerts a  downward forcing weight of $-3.328$. Because the intermediate slow modes possess their own deep temporal history, this energy cascade is highly structured and non-Markovian. To decode this intricate temporal footprint being handed down across the boundary, the PLS algorithm extracts 13 latent components from the 23 lagged cross-coupled predictors available to the fast group. This allows the model to deterministically organize the formation of individual clouds based on the cascading inertia-gravity wave dynamics while conservatively compressing the raw dimensional space.

In the context of the PLS decomposition, the deterministic weights represent the structured, predictable forcing derived explicitly from the resolved variables (the instantaneous cascades and the specific memory lags). If these weights experience a relative dip, it does not mean the modes are pure noise; it just means the model is loosening its rigid structural constraints, allowing the unresolved, chaotic physics (the stochastic noise term) to play a proportionally larger role in that specific spectral window.

This delicate balance between structure and stochasticity is vividly captured by the distinct U-shaped ``bathtub'' profile across the fast-scale spectrum. At the spectral boundaries ($\text{PC}_{11}$ and $\text{PC}_{16}$), the modes are subjected to deterministic forcing---representing the highly structured injection of mesoscale energy at the top and the rigidly organized dissipation of mixing processes at the bottom. In contrast, the intermediate convective modes ($\text{PC}_{12}$ through $\text{PC}_{15}$) exhibit exactly the type of deterministic dip described above. In this middle zone, the structured predictability is comparatively relaxed, allowing the stochastic noise innovations to step in and drive the variance. Physically, this structural relaxation mirrors the inherent unpredictability of the boundary layer, where the intermediate lifecycle stages of  convective updrafts are highly transient and governed by turbulent turnover.

At the opposite end of the resolved spectrum ($\text{PC}_{16}$), the total deterministic volume spikes massively, driven overwhelmingly by the short-term memory ($\tau_{\mathrm{fst}_1}  = 0.16$~h). This roughly 10-minute lag captures the immediate local autocorrelation of the convective pulses, physically acting as a proxy for rapid entrainment and mixing. Because these turbulent exchanges operate much faster than broader boundary layer recirculation, they impose a heavy autoregressive penalty (reaching $-5.295$ for $\text{PC}_{16}$). This mathematical penalty effectively parameterizes the rapid dissipation of the smallest individual clouds, preserving the system's structural stability. Conversely, the deep memory ($\tau_{\mathrm{fst}_2}  = 2.54$~h) visually anchors the deterministic forcing at both spectral boundaries ($\text{PC}_{11}$ and $\text{PC}_{16}$). This 2.5-hour timescale confirms that localized cloud fluctuations do not emerge spontaneously; rather, their rapid formation is structurally preconditioned by the broader mesoscale organization and slowly evolving meteorological background established hours prior \cite{dror2022uncovering}.

\begin{figure}[htbp]
    \centering
    \includegraphics[width=1\linewidth, height=0.5\textwidth]{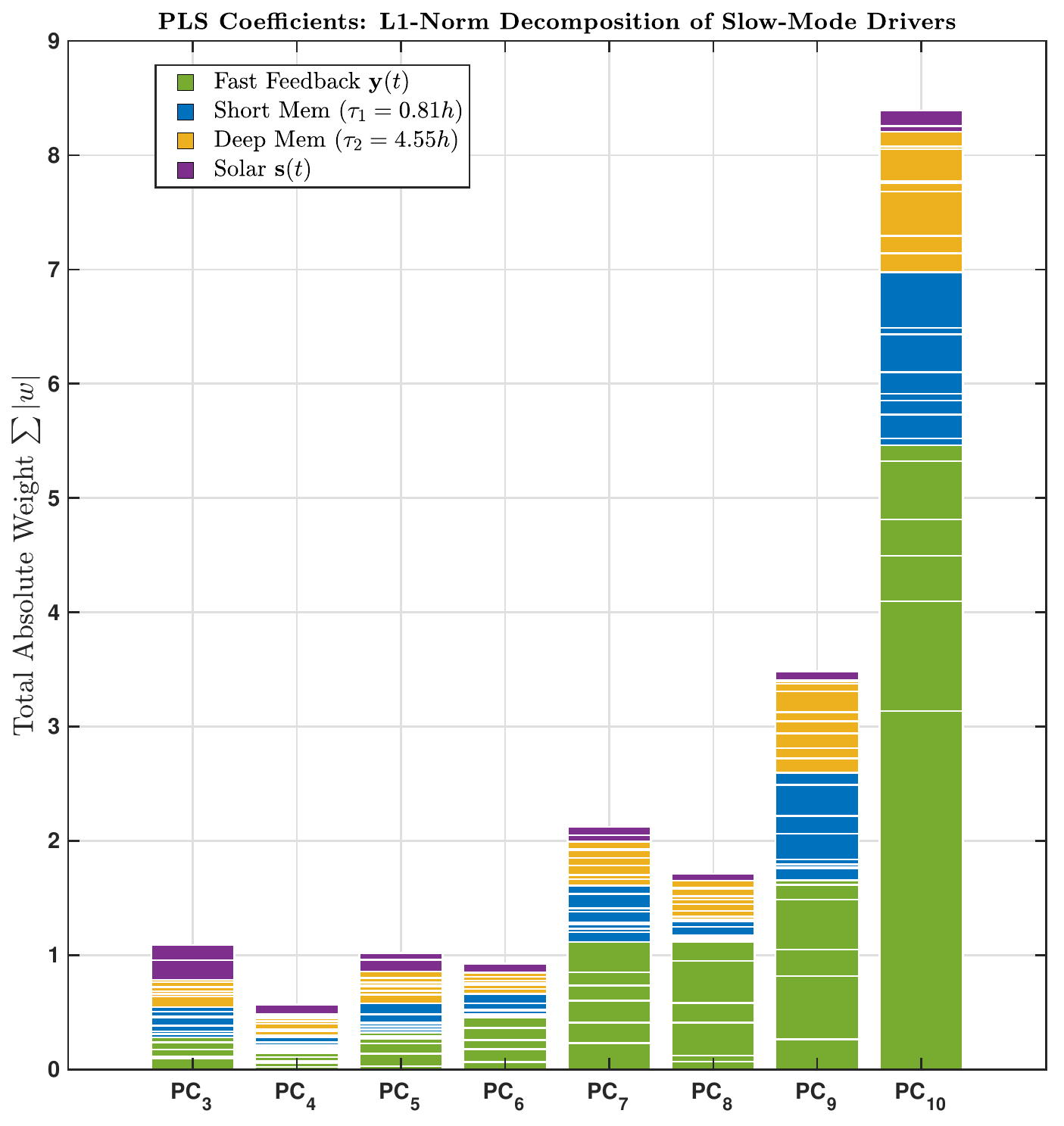}
    \caption{\textbf{$L_1$-Norm Decomposition of Slow-Mode Predictors.} Total absolute deterministic weight ($\sum |w|$) applied to the mesoscale slow-mode equations ($\text{PC}_{3}$--$\text{PC}_{10}$). Predictors are categorized into: high-frequency backscatter from the localized convective fast modes (green), gravity-wave phase shift memory at $\tau_{\mathrm{slw}_1}  = 0.81$~h (blue), deep spatial diurnal memory at $\tau_{\mathrm{slw}_2}  = 4.55$~h (yellow), and solar forcing (purple). The massive feedback volume at $\text{PC}_{10}$ indicates intense, bidirectional spectral boundary interactions, while the low-volume core ($\text{PC}_{3}$--$\text{PC}_{6}$) identifies persistent, autonomous inertia-gravity wave propagation.}
    \label{PLS_L1-norm_slow}
\end{figure}

Turning our attention to the intermediate-frequency slow modes (Fig.~\ref{PLS_L1-norm_slow}), the decomposition uncovers the intrinsic memory of the mesoscale environment and its bidirectional relationship with the localized convective scales. To manage this complexity, the slow group extracts 11 PLS components from a  pool of 25 lagged cross-coupled predictors to forecast its 8 variables. The most striking structural feature of the slow group's forcing is the ``wall of feedback'' at $\text{PC}_{10}$. Its total deterministic weight spikes dramatically and is almost entirely composed of the fast feedback mechanism ($\mathbf{y}(t)$). Sitting directly on the spectral boundary, this  upward forcing proves that energy transfer is not a one-way mechanism. While the mesoscale organization cascades energy down, the rapid, localized fluctuations of the individual clouds are simultaneously backscattering structural information up into $\text{PC}_{10}$, actively disrupting and shaping the smallest of the inertia-gravity wave scales.

Furthermore, the lower-frequency modes of the slow group ($\text{PC}_3$ through $\text{PC}_6$) exhibit a significantly lower total deterministic weight compared to the rest of the group. While the absolute solar contribution remains relatively constant across all slow modes (and is slightly higher for $\text{PC}_3$), it becomes proportionally more dominant for these lower-frequency  modes due to their smaller total weight. Consequently, within $\text{PC}_3$ through $\text{PC}_6$, the solar contribution is more evenly balanced proportionally by the influence of deep memory ($\tau_{\mathrm{slw}_2} $) and the feedback of the fast variable $\mathbf{y}(t)$---even though the absolute contribution of the latter is small in this subgroup, implying that these modes are mainly persistent and autonomous. As we will establish in the subsequent spectral analysis, these specific pairs represent higher-order diurnal sub-harmonics (such as the quarter-diurnal cycle).

 The model elegantly captures the transition from these persistent sub-harmonics to the faster inertia-gravity waves ($\text{PC}_7$ through $\text{PC}_{10}$) through its algorithmic selection of the two slow memory lags. The first lag, $\tau_{\mathrm{slw}_1}= 0.81$~h, specifically tunes the propagation of the mesoscale gravity waves.  As detailed in Table~\ref{Table_Predicted_Freqs} below, the highest-frequency  pair in the slow block ($\text{PC}_{9}-\text{PC}_{10}$) exhibits a dominant period of $\sim 1.5$~hours. The selected lag of $0.81$~h is about half the period of this wave. Mathematically, lagging an oscillator by half its period allows the delayed system to explicitly encode phase shifts, which is essential for simulating structural wave propagation across the spatial domain.

 Finally, the broad spatial environment is governed by the longest delay, $\tau_{\mathrm{slw}_2} = 4.55$~h. While the mesoscale gravity waves rely on the intermediate phase-shift lag ($\tau_{\mathrm{slw}_1}$), the broader sub-harmonic oscillations ($\text{PC}_3$ through $\text{PC}_6$) are structurally anchored to this much deeper 4.5-hour foundation. This delay correlates directly with the extended diurnal heating cycle and the vast spatial organization of the cumulus fields over the landmass. Recent machine-learning classification analyses of extensive satellite datasets have established that the formation of these continental GreenCu fields is universally governed by specific large-scale synoptic meteorology, such as broad regional subsidence and lower-tropospheric stability \cite{dror2022uncovering}. Consistent with the MZ closure interpretation discussed in {\it Methods: Concatenated Time and the Role of Delay-Memory}, this deep 4.5-hour delay physically acts as the slowly varying deterministic baseline that parameterizes these exact synoptic constraints. By capturing the unobserved thermodynamic inertia of the overarching subsidence, it effectively preconditions the large-scale meteorological macro-environment long before the rapid, localized boundary layer overturning takes place.


\subsection*{Out-of-Sample Spectral Recovery and Emergent Physical Regimes}
Having established the structural and physical integrity of the MSHDM's temporal decomposition, we now evaluate the framework's generative capabilities. The ultimate test of a stochastic emulator lies not in its ability to regress against known training data, but in its capacity to generate long-term, out-of-sample (OOS) trajectories that remain statistically and dynamically consistent with the underlying complex system. As highlighted in the core thesis of this work, accurately capturing the emergent, multiscale beat patterns of continental cloud fields requires overcoming the inherent spectral sparsity of standard finite-dimensional stochastic models. 

\begin{figure}[htbp]
    \centering
    \includegraphics[width=1\linewidth, height=0.6\textwidth]{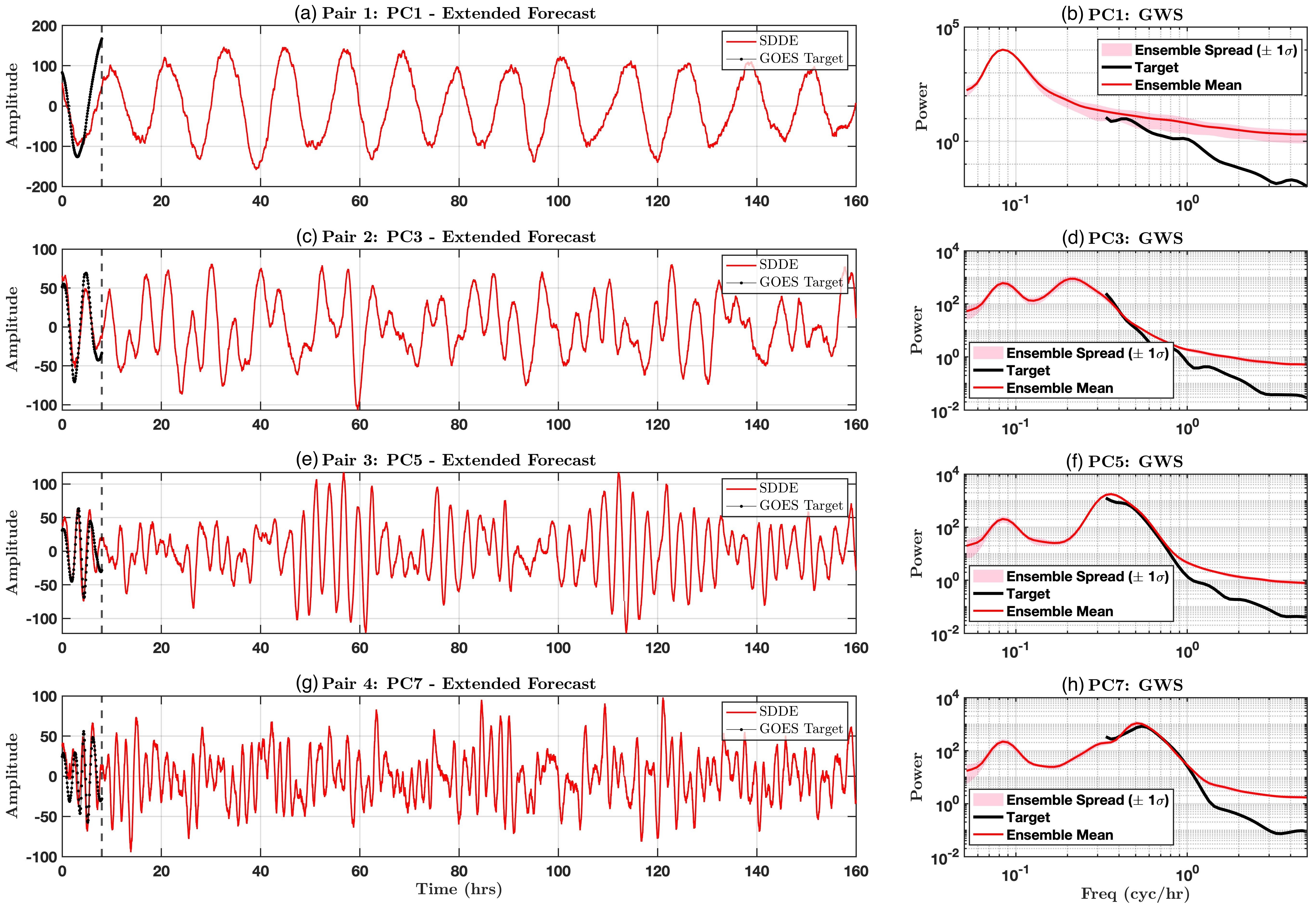}
    \caption{\textbf{Time Series and GWS for Pairs 1--4.} The left panels show a selected realization of the extended stochastic integration (red) compared against the true GOES-16 trajectory (black dotted). The right panels display the out-of-sample spectral recovery. The solid red line denotes the mean GWS of the top 100 OOS ensemble realizations (selected via iterative variance calibration), surrounded by the $1\sigma$ ensemble spread (shaded pink). The emulator successfully tracks the empirical target GWS (solid black), matching both the dominant low-frequency solar harmonic in Pair 1 and the transition into broader intermediate-frequency dynamics.}
    \label{Fig_Dash_1to4}
\end{figure}

\begin{figure}[htbp]
    \centering
    \includegraphics[width=1\linewidth, height=0.5\textwidth]{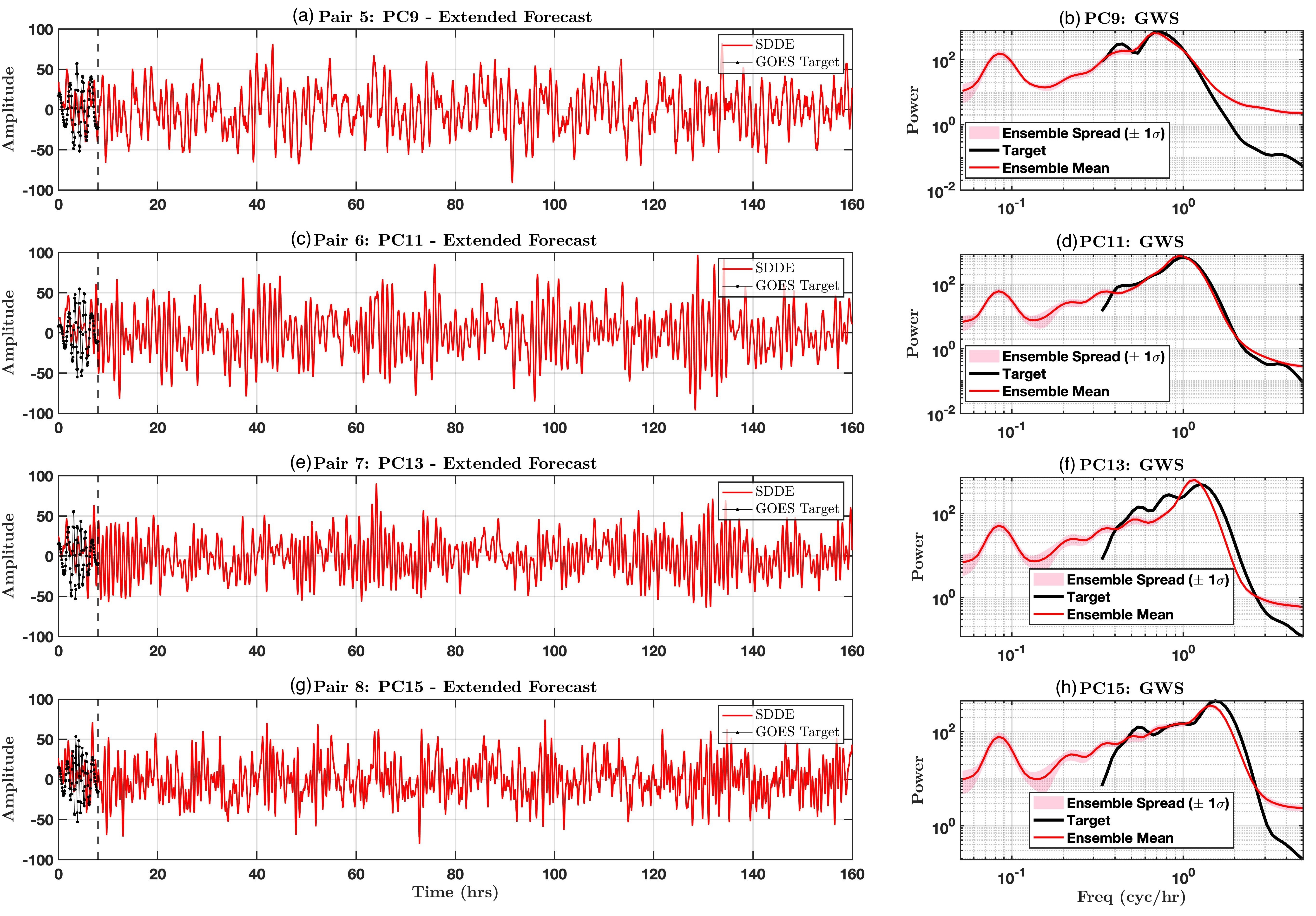}
    \caption{\textbf{Time Series and GWS for Pairs 5--8.} Extended forecasts and spectral validation for the deeper, higher-frequency modes governing localized convective lifecycles. Despite the highly transient nature of these boundary layer scales, the MSHDM effectively preserves the broadband spectral shape and limits variance drift. The high-frequency GWS tail exhibits significantly better agreement in these deeper modes, reflecting the higher sampling density of convective events within the training window.}
    \label{Fig_Dash_5to8}
    \vspace{-.1cm}
\end{figure}

To validate this point, we compute an ensemble of OOS forecasts (utilizing the optimally calibrated pruned architecture) to ensure our results reflect the true structural stability of the learned attractor rather than a single fortuitous noise realization. The 100 best-performing members are selected via the Iterative Calibration of Stochastic Forcing algorithm (see {\it Methods}), which strictly enforces the fluctuation-dissipation balance to prevent energy collapse. The extended forecasting results for the leading topological pairs are presented in Fig.~\ref{Fig_Dash_1to4} and Fig.~\ref{Fig_Dash_5to8}. The left panels illustrate the simulated time series operating in a purely autonomous OOS forecast regime spanning approximately 160 hours. The right panels display the corresponding GWS, providing a definitive frequency-domain validation by comparing the true empirical target (solid black) against the 100-member OOS ensemble mean (solid red) and its $1\sigma$ variance spread (shaded pink).

\begin{table}[htbp]
    \centering
\caption{\textbf{MSHDM Predicted Peak Frequencies.} Listed are the peak frequencies generated natively by the MSHDM out-of-sample 160-hour extended simulation. These predicted frequencies are extracted directly from the peak energy of the simulated GWS ensemble mean (red curves in the right panels of Figs.~\ref{Fig_Dash_1to4}--\ref{Fig_Dash_5to8}). As discussed in the main text, the model resolves the semidiurnal forcing and its mesoscale sub-harmonics (Pairs 1--3) in the low-frequency range, which remains inaccessible in the observations due to the short 8-hour record. Periods exceeding 120 minutes are displayed in hours. These predicted frequencies dictate the placement of the white dashed lines in the accompanying scalograms (Figs.~\ref{Fig_Scalo_1to4}--\ref{Fig_Scalo_5to8}). Notably, for the resolvable frequencies, these MSHDM predictions fall within a tight $\sim 5\%$ error margin compared to the empirical observational targets.}
    \vspace{0.15cm}
    {\footnotesize
    \begin{tabular}{l c c c}
        \toprule
        \textbf{Group} & \textbf{Conjugate Pair} & \textbf{Predicted Freq (cyc/hr)} & \textbf{Predicted Period} \\
        \midrule
        \textbf{Solar}  
        & Pair 1 ($\text{PC}_{1}$--$\text{PC}_{2}$) & 0.0837 & 11.95 hrs \\
        \midrule
        \multirow{4}{*}{\textbf{Slow}} 
        & Pair 2 ($\text{PC}_{3}$--$\text{PC}_{4}$) & 0.2062 &  4.85 hrs \\
        & Pair 3 ($\text{PC}_{5}$--$\text{PC}_{6}$) & 0.3590 &  2.79 hrs \\
        & Pair 4 ($\text{PC}_{7}$--$\text{PC}_{8}$) & 0.5077 & 118.19 min \\
        & Pair 5 ($\text{PC}_{9}$--$\text{PC}_{10}$) & 0.6699 & 89.57 min \\ 
        \midrule
        \multirow{3}{*}{\textbf{Fast}} 
        & Pair 6 ($\text{PC}_{11}$--$\text{PC}_{12}$) & 0.9473 & 63.34 min \\
        & Pair 7 ($\text{PC}_{13}$--$\text{PC}_{14}$) & 1.1663 & 51.45 min \\
        & Pair 8 ($\text{PC}_{15}$--$\text{PC}_{16}$) & 1.4359 & 41.79 min \\
        \bottomrule
    \end{tabular}
    }
    \label{Table_Predicted_Freqs}
\end{table}
A key achievement of the MSHDM framework is the fidelity of this out-of-sample spectral recovery. Across the hierarchy of modes, the tight $1\sigma$ ensemble spread explicitly demonstrates that the model maintains long-term structural stability without suffering from variance inflation. The ensemble mean curves (solid red) track the peak energy maxima of the target GWS  (black curve) with a good level of accuracy, successfully recovering the broad-band peaks that define the system's principal variance. 

Another key achievement of the extended simulations is their successful prediction beyond the empirical resolution limit imposed by the 8-hour observational window for Pairs 2 and 3 in the slow modes. For these pairs, the model natively predicts GWS peaks with periods of $4.85$~hours and $2.79$~hours, respectively; see Table \ref{Table_Predicted_Freqs}. This, combined with the $11.95$-hour GWS peak for Pair 1, provides a powerful physical interpretation of the $\text{PC}_1$--$\text{PC}_6$ subgroup that is otherwise inextricable from pure observations. 

We can indeed state that the model natively generates dominant periods for Pairs 2 and 3 that are consistent with the higher-order quarter-diurnal and eighth-diurnal sub-harmonics (Fig.~\ref{Fig_Dash_1to4}). In boundary layer meteorology, the diurnal cycle of continental shallow cumulus is highly asymmetric. As demonstrated using an ensemble of large-eddy simulations \cite{brown2002large}, this lifecycle is characterized by the breakdown of the nocturnal inversion by morning sensible heating, triggering a rapid onset of cloud cover that peaks early in the growth phase, followed by a sharp evening collapse as surface forcing decays. This non-linear lifecycle introduces sharp, transient edges into the daily reflectance signal. It is remarkable that the MSHDM enables us to reveal how the EOF decomposition partitions this asymmetric variance into higher-order temporal sub-harmonics, manifested by the quarter-diurnal and eighth-diurnal cycles emerging directly as the principal components of Pairs 2 and 3. These serve as the fundamental structural basis functions required to resolve the rapid diurnal onset and collapse.


 It is important to acknowledge that a mismatch in the high-frequency tail of the GWS is visible for the leading macroscopic pairs, which is expected given the large statistical uncertainty associated with high-frequency variance estimates derived from a severely limited 8-hour observational dataset. However, this discrepancy is noticeably reduced as we progress deeper into the spectrum (e.g., Pairs 5--8), where the high-frequency convective events are more densely sampled within the training window.

    \vspace{-.1cm}

%

Physically, the frequencies generated by the stochastic ensemble (Table \ref{Table_Predicted_Freqs}) map comprehensively across the known physical regimes of boundary layer and mesoscale dynamics, bridging the synoptic scales down to the turbulent convective engine. 

Moving into Pairs 4 and 5 (118.19 min and 89.57 min), the emulator resolves the mesoscale organization and inertia-gravity waves that separate synoptic forcing from localized turbulence. This 1.5 to 2-hour band perfectly brackets the near-periodic 1.5-hour oscillation previously identified in Hovmöller space as the empirical fingerprint of convective IGWs traversing this specific continental cloud field (misattributed to the $\text{PC}_5$--$\text{PC}_6$ pair in \cite{Dror2021}). As established in fundamental numerical studies, these mesoscale waves are primarily excited by convective overshooting pushing into the stable stratification above, and boundary layer eddies acting as physical obstacles to the environmental wind shear \cite{clark1986convectively, lane2015gravity}. Once generated, they actively modify the local thermodynamic environment, altering convective inhibition and triggering low-level lifting \cite{lane2001gravity}. The simulated reflectance successfully captures this dynamic boundary-layer feedback, oscillating organically as the propagating waves create localized, organized bands of uplift (condensation) and subsidence (clearing) \cite{lane2002gravity, Dror2021}.

Finally, delving into the convective lifecycles, the high-frequency cluster generated by the Fast modes (Pairs 6 through 8, scaling from $\sim 64$ min down to $\sim 42$ min) represents the turbulent, convective engine of the cloud field. To physically interpret these specific periodicities, we must apply dimensional scaling to the established dynamics of shallow cumulus lifecycles. High-resolution tracking of marine shallow cumulus demonstrates that individual thermals possess a characteristic turnover time, $t_* = H / w_{up}$, of approximately 10 minutes. In this formulation, $H$ defines the macroscopic vertical length scale of the system (the cloud depth), and $w_{up}$ represents the characteristic updraft velocity. The complete cloud lifecycle---spanning multiple ascending pulses and eventual evaporative dissipation---typically lasts roughly 2.5 times this turnover period \cite{zhao2005life_part1, zhao2005life_part2}. For continental GreenCu fields, the convective boundary layer and resulting cloud depths ($H$) are roughly 1.5 to 2 times deeper than their marine counterparts \cite{brown2002large}. Scaling $t_*$ proportionally to these continental dimensions yields an extended thermal turnover time of 15 to 25 minutes. Consequently, the complete multi-pulse lifecycle scales directly into a 40 to 60-minute band. The 63.34 min and 51.45 min peaks elegantly capture this full, extended continental lifecycle. Conversely, the faster 41.79 min peak reflects the period of the individual, high-frequency updraft pulses (or thermals) driving the ascending cloud top before mixing and collapsing \cite{zhao2005life_part2}. The successful spectral recovery of these rapid modes perfectly aligns with previous spatiotemporal analyses of this dataset, which demonstrated that these exact high-frequency EOFs capture the individual clouds that continuously form and dissipate like ``pearls on a string'' along the broader mesoscale cloud streets; see Fig.~6f in \cite{Dror2021}.

\vspace{-.1cm}
\begin{figure}[htbp]
    \centering
       \includegraphics[width=1\linewidth, height=0.55\textwidth]{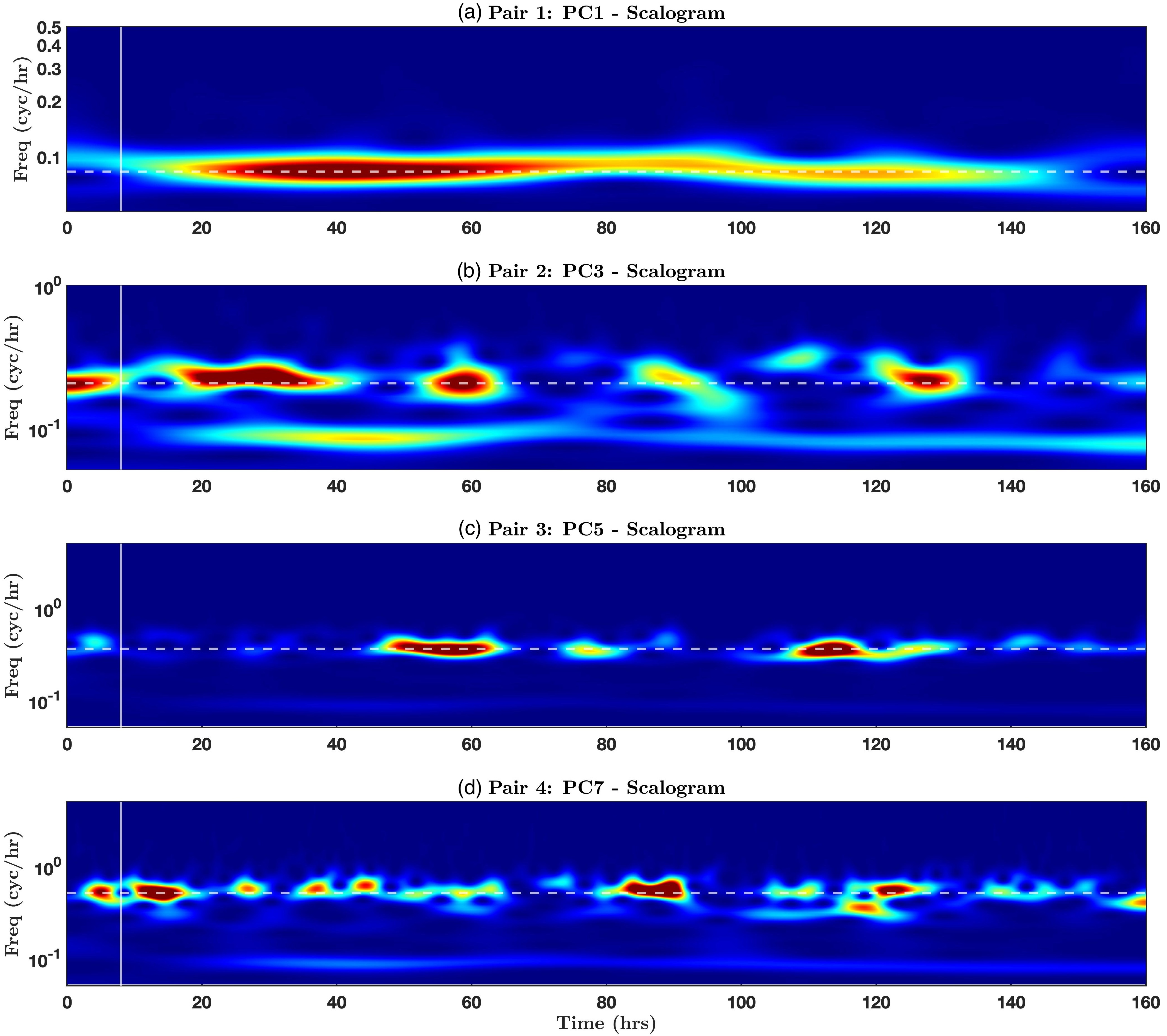}
    \caption{\textbf{Out-of-Sample Scalogram Forecasts for Pairs 1--4.} Time-frequency representations of the simulated trajectories. The horizontal white dashed lines indicate the predicted peak frequencies generated natively by the model (detailed in Table~\ref{Table_Predicted_Freqs}). These scalograms show the model's ability to resolve the semidiurnal cycle and its mesoscale sub-harmonics (Pairs 1--3) beyond observations. While the diurnal band of Pair 1 remains relatively stable over time (due to its strong periodic component, see top-left panel in Fig.~\ref{Fig_Dash_1to4}), Pairs 2 to 4 present localized energetic bursts that gently wander around these predicted baselines. This temporal localization and spectral wandering are physical fingerprints of emergent AM-FM dynamics. Note that Pair 2 retains a faint trace of the underlying limit cycle, consistent with its quarter-diurnal nature being more heavily influenced by the fundamental solar envelope.}
    \label{Fig_Scalo_1to4}
    \vspace{-.1cm}
\end{figure}
\vspace{-.1cm}
\begin{figure}[htbp]
    \centering
    \includegraphics[width=1\linewidth, height=0.55\textwidth]{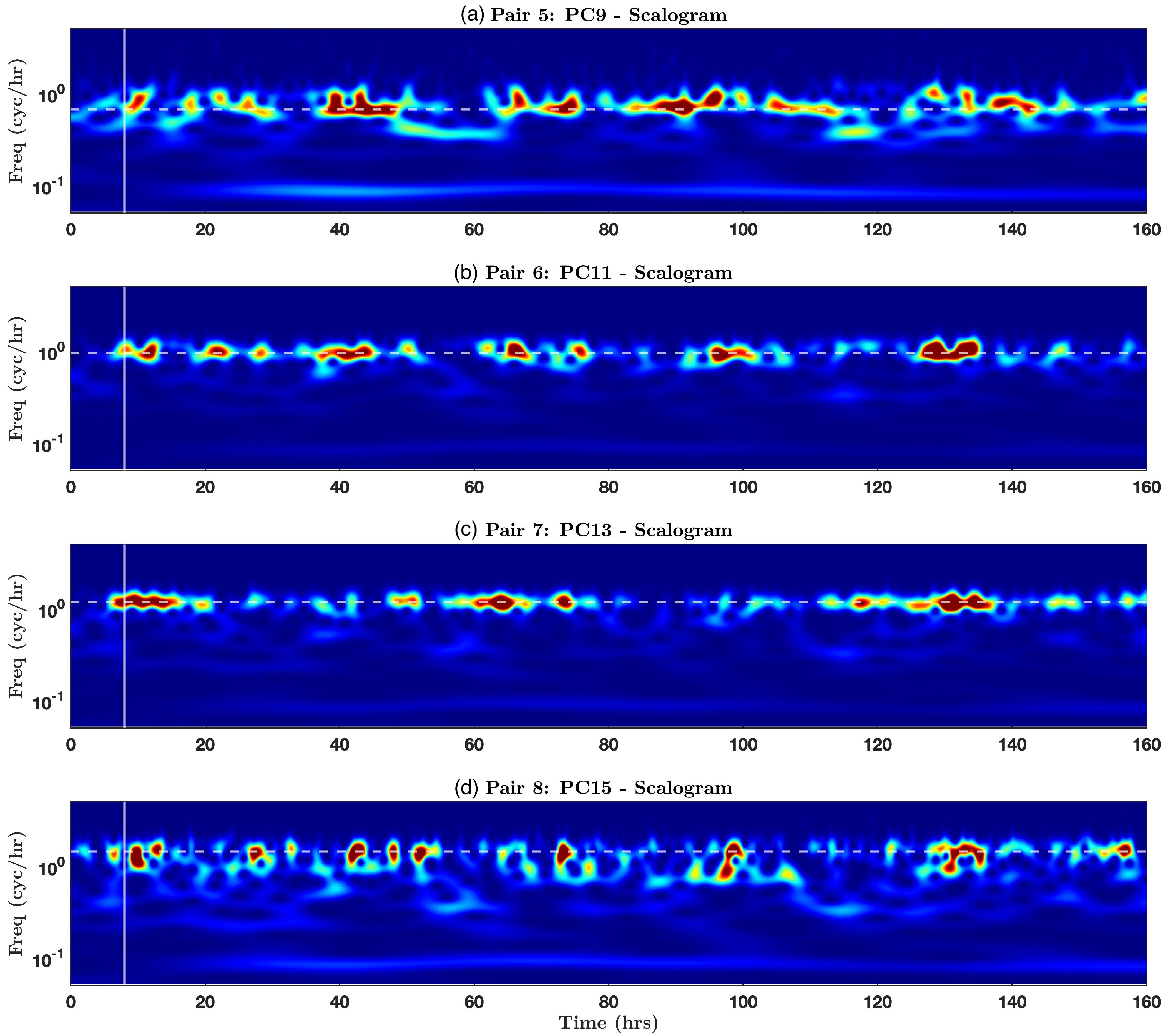}
    \caption{\textbf{Out-of-Sample Scalogram Forecasts for Pairs 5--8.} Time-frequency representations for the localized convective scales. As in Fig.~\ref{Fig_Scalo_1to4}, the horizontal white dashed lines denote the predicted peak frequencies for these pairs (detailed in Table~\ref{Table_Predicted_Freqs}). The ``puffing'' AM-FM beating becomes increasingly pronounced at these higher frequencies, successfully mirroring the highly transient, stochastic lifecycles of localized convective cloud formations (see \textit{Methods: Cloud Field Simulation: Phenomenological Analysis} and the \textit{Supplementary Movie}).}
    \label{Fig_Scalo_5to8}
    \vspace{-.1cm}
\end{figure}

While the GWS confirms that the time-averaged spectral energy is conserved by the ensemble, it is the transient, time-localized behavior that truly defines the complexity of the cloud field. To verify that the MSHDM is generating physically realistic, non-stationary wave packets, we project the simulated OOS trajectories into the time-frequency domain using continuous wavelet transforms. 

The resulting scalograms (Fig.~\ref{Fig_Scalo_1to4} and Fig.~\ref{Fig_Scalo_5to8}) vividly expose the emergence of complex, "puffing" beat patterns. Rather than presenting as solid, continuous horizontal bands of uniform power, the simulated energy is characterized by highly localized, intermittent bursts. This temporal localization, combined with wandering frequencies that gently drift around the target peaks, is the precise dynamical signature of coupled AM-FM dynamics. A notable exception is the faint continuous strip observed below the transient bursts in Pair 2 (Fig.~\ref{Fig_Scalo_1to4}); this trace of the underlying limit cycle is physically consistent with its status as a quarter-diurnal mode, which is more heavily anchored to the fundamental solar envelope than the deeper convective pairs, supported by the corresponding GWS displaying a $12$-hour bump of similar magnitude to the $5$-hour peak. 

The MSHDM does not enforce these AM-FM beat patterns through explicit periodic forcing; instead, they emerge organically from the interplay of the deep convective memory lags, the forward energy cascade, and the calibrated stochastic noise. This specific interplay---convective inertia-gravity waves organizing localized convective lifecycles into distinct, transient strips---is a pattern typically observed in cumulus convection over land \cite{plank1966wind, plank1969size, da2011cloud, Dror2021}. The spatio-temporal scales  matched by the emulator are highly consistent (see {\it Supplementary Movie}) with those previously documented in the broader literature for shallow cumulus clouds and atmospheric gravity waves \cite{plank1969size,  lane2015gravity, Dror2021}. This confirms that the framework successfully captures the true transient, non-stationary life cycles of the cloud deck, establishing the MSHDM as a robust and efficient modeling framework for generating complex AM-FM dynamics from limited observational data.


\medskip

\section*{Discussion}
 Extracting coherent, cross-scale dynamics from short observational records exposes a fundamental bottleneck in standard finite-dimensional stochastic modeling: the inability to generate sufficient spectral density without resorting to massive, heavily parameterized equations. The Multilayered Stochastic Hierarchical Delay Model (MSHDM) provides a rigorous mathematical solution to this constraint. By formulating the governing dynamics through a hierarchy of continuous fractional memory lags, the framework elevates the system into an infinite-dimensional phase space, naturally yielding the dense, highly modulated spectral signatures required for complex systems. However, calibrating such an intricately delayed structure directly from transient data requires a departure from conventional fitting techniques. To achieve this, we developed a hybrid offline-online Bayesian optimization strategy that abandons standard pointwise trajectory matching---an approach that typically fails to capture coherent wave properties in chaotic systems. Instead, the algorithm autonomously determines the optimal latent Partial Least Squares (PLS) basis and continuous temporal delays by enforcing strict structural and energy-cascade fidelity against the empirical Global Wavelet Spectrum (GWS).

Applied to high-resolution satellite observations of continental cloud fields, the resulting MSHDM (Eqns.~\eqref{eq:sdde_solar}-\eqref{eq:sdde_fast}) functions as a powerful stochastic pattern generator for spatiotemporal evolution. By generating a large, statistically diverse ensemble, the framework successfully bridges three distinct dynamical regimes: the macroscopic semidiurnal harmonics and atmospheric tides, the intermediate mesoscale organization driven by inertia-gravity waves, and the rapid turbulent turnover of individual convective lifecycles. Unifying these scales allowed us to extract dynamical insights and structural couplings that were statistically latent and otherwise obscured in the raw, short satellite record. Furthermore, the out-of-sample matching of continuous wavelet scalograms, complemented by the explicit quantification of the underlying cross-scale physical drivers via the $L_1$-norm PLS decomposition, confirms the model's remarkable skill in natively generating the true transient, non-stationary evolution of the cloud deck. 
The physical manifestation of these coupled dynamics is most vividly corroborated by the extended out-of-sample cloud field simulation provided in the Supplementary Movie discussed in {\it Methods: Cloud Field Simulation: Phenomenological Analysis}. As observed, the MSHDM successfully sustains multiscale phase coherence across multiple asymmetric diurnal cycles (those shown in top left-panel of Fig.~\ref{Fig_Dash_1to4}), organically generating the complex AM-FM beat patterns without succumbing to numerical diffusion.

A key dynamical insight is that although each Principal Component (PC) exhibits complex Amplitude and Frequency Modulation (AM-FM), the PCs also display a clear frequency-ranking property by their mid-day dominant frequency. This contrasts sharply with highly non-linear systems, such as rotating stratified turbulent flows, where PCs typically display a complex mixture of slow and fast frequencies \cite{LucariniChekroun2023}. In such cases, the data-driven modeling of coarse-grained oceanic turbulence necessitates highly non-linear, coupled Stuart-Landau systems to properly emulate complex dynamics like jet-eddy interactions \cite{KCB18}.

The successful deployment of a linear MSHDM to represent these complex cloud field dynamics naturally raises the question of nonlinearity in the underlying physics. While it is not guaranteed that the organization of this specific shallow convective cloud field is strongly nonlinear, the successful calibration of a linear model from a short observational burst relies on two fundamental points of structural compensation.

The first point of compensation lies in the use of the EOF subspace: projecting the high-dimensional nonlinear dynamics onto a truncated, low-dimensional latent space constitutes a change of coordinates that can effectively linearize or regularize complex behaviors within that specific subspace \cite{brunton2017chaos}. The second, and more fundamental, mechanism is the MSHDM structure itself. Formulated as a type of Generalized Langevin Equation, the model's delayed, multilevel linear form is mathematically equipped to compensate for residual nonlinearities. As explicitly quantified by our $L_1$-norm decomposition of the PLS driving weights, the system's continuous fractional memory and the cross-layer coupling between the slow and fast turbulent modes ($\mathbf{x}$ and $\mathbf{y}$ layers) collectively act as a highly efficient \textit{effective linear closure}. This closure natively represents the nonlinear energy cascade and unresolved turbulent backscatter in the latent space \cite{LucariniChekroun2023}. Alternatively, the turbulent subsystems can be viewed as a linear network forced by the overarching solar cycle ($\mathbf{s}$-layer) that successfully recovers nonlinear dynamics, consistent with the findings of \cite{brunton2017chaos}.

The ability of the MSHDM to function as a computationally efficient, high-fidelity stochastic emulator opens critical new avenues for fundamental climate research. Rather than relying exclusively on massive empirical dataset classifications to uncover universal convective behaviors \cite{dror2022uncovering}, the MSHDM framework provides a powerful alternative approach. By extracting an analytical stochastic dynamical model directly from a single, short observational record, the framework allows us to explicitly interrogate the underlying physical mechanisms at play. Ultimately, because we gain a low-dimensional, data-driven structured model---and not merely a ``black-box'' statistical data generator---we can directly decode the cross-scale energy transfers governing the cloud field. As demonstrated, this physical inference is achieved by analyzing the model's calibrated internal components, specifically through the $L_1$-norm decomposition of the PLS driving weights (Figs.~\ref{PLS_L1-norm_fast} and \ref{PLS_L1-norm_slow}) and the GWS analysis of the emergent multiscale dynamics (Figs.~\ref{Fig_Dash_1to4} and \ref{Fig_Dash_5to8}).

The MSHDM's interpretable attributes offer a promising framework to address the unsolved question of what large-scale meteorological conditions allow GreenCu to prevail globally \cite{dror2022uncovering}. Previous work established that GreenCu fields share similar characteristics across diverse geographical locations, dictated by distinct profiles of temperature, humidity, and strong free-tropospheric subsidence \cite{dror2022uncovering}. We propose moving beyond simple composite analyses of these mean vertical profiles. The next logical step is to train \textit{multivariate MSHDMs} that directly incorporate these large-scale meteorological drivers. This would involve learning models trained not only on the GreenCu visible-range PCs, but also on PCs derived from co-located fields of potential temperature ($\theta$), specific humidity ($q$), and relative humidity ($\textrm{RH}$) extracted from reanalysis data (e.g., ERA5 \cite{hersbach_2020era5}). By systematically learning MSHDMs for cloud events across different geographical and climatic regimes, one could isolate the dynamic, coupled signatures in the reduced phase space that dictate shallow cloud formation. The knowledge gained from this approach would provide valuable insights for the parameterization of shallow cumulus in General Circulation Models (GCMs) and significantly advance our understanding of low-cloud feedback on climate sensitivity. Naturally, this application extends to marine stratocumulus clouds exhibiting other types of coherent structures, such as open and closed cells.

Finally, the success of the MSHDM framework in capturing the complex, coupled AM-FM dynamics of cloud organization highlights the methodology's broad, universal utility. A key mathematical distinction of the MSHDM lies in the time-invariant nature of its coefficients. In traditional modeling approaches, complex AM-FM dynamics are typically produced by explicitly including relevant deterministic, time-dependent forcing functions into the model. By contrast, the MSHDM generates these complex modulations \textit{autonomously} through its high spectral density and layered, continuous lag structure, bypassing the need to hardcode deterministic time-dependent forcing.

By providing a mathematically rigorous, hybrid data-driven framework capable of resolving intricate spectral structures and their coupled intensity-frequency variations, the MSHDM overcomes the critical limitations of standard finite-dimensional models. It serves as a highly compressible, general-purpose tool for resolving latent AM-FM beat patterns, offering a novel interpretability framework through hierarchical delays. Ultimately, this structural adaptability positions the MSHDM to decode emergent wave-packet phenomena across a vast spectrum of complex systems---scaling seamlessly from the microscopic dynamics of biological systems and engineered optoelectronics \cite{al2018multifrequency, buzsaki2004neuronal, soriano2013complex} to the macroscopic scales of geophysical turbulence \cite{LucariniChekroun2023} and astrophysical gravitational waves \cite{abbott2016observation}.

{\small

%
\section*{Methods}
\subsection*{GOES--16 ABI dataset and its preprocessing}
\noindent  The dataset used in the present study is the same as analyzed in \cite{Dror2021}. We summarize below some elements about this dataset and refer to  \cite{Dror2021} for more details. 
The Advanced Baseline Imager (ABI) on board GOES--16 is a state-of-the-art 16-band radiometer, providing four times the spatial resolution, and more than five times faster temporal coverage than the former GOES system \cite{schmit_2017}. We use the ABI's level 1B ``Red'' band (channel 2, $0.64\,\mu m$) radiance, which has the finest resolution (0.5 km) of all ABI bands, over continental US  (temporal resolution of 5 minutes). The "Red'' band detects reflected visible solar radiation, and its high-resolution makes it ideal for exploring greenCu during daytime.  To prepare this dataset for EOF decomposition, we converted the radiance values to reflectance following \cite{schmit_2010}, applied a simple gamma correction to adjust and brighten the images (see below), and reprojected the data from their native geostationary projection into a geographic (latitude--longitude) one. We focus on a vast region of interest (ROI), located between $30^{\circ}N-36^{\circ}N$ and $95^{\circ}W-80^{\circ}W$, on a day that features a typical case of daytime, locally formed shallow convection, from late-morning (10:47 Eastern standard time [EST]) to the afternoon (18:47 EST) of August 22nd, 2018, corresponding to a scalar field comprised of 97, 2D snapshots of corrected reflectance over $1335$ latitude grid points ($\phi$) and 3339 longitude grid points ($\lambda$), for a total of eight hours. The ROI spans two different time zones, such that the local time at the eastern and western parts are four and five hours behind coordinated universal time (UTC), respectively.


\subsection*{Gamma correction}
\noindent   The reflectance images obtained from GOES--16 ABI's ``Red'' band looked somewhat dark, since the values were in linear units. We adopted thus the common practice which consists of adjusting and brightening the images (before performing the EOF decomposition  \cite{Dror2021}) by applying a simple gamma correction as follows:
$I_{out}=AI_{in}^\gamma$, where $I_{in}$ is the darker input image, $\gamma=0.5$, $A=1$, and $I_{out}$ is the adjusted image (corrected reflectance).

%

\subsection*{Concatenated Time and the Role of Delay-Memory}
\noindent Because the observational reflectance dataset is intrinsically constrained to daylight hours, the temporal coordinate $t$ in the SDDE formulation is defined as concatenated daylight, bypassing the unobserved nocturnal regime. Thus, the time axis $t$ must be interpreted strictly as cumulative daylight hours, where $t$ in $[0,12)$ is Day 1, $t$ in $[12,24)$ is Day 2, and so forth. This way, the model natively stitches these roughly 12-hour daylight windows end-to-end, aligning its fundamental mathematical cycle directly with the satellite's observational window. 

To account for the severe state-space reduction---from the full 3D atmospheric boundary layer down to the reduced phase space  of selected PCs from the visible-range (reflectance field)---we rely on the MZ projection formalism discussed previously. Because our current emulator is trained exclusively on the visible-range PCs of the GreenCu field, it is physically ``blind'' to the fundamental atmospheric drivers that dictate continental shallow cloud formation, such as potential temperature ($\theta$), specific humidity ($q$), relative humidity ($\text{RH}$), and free-tropospheric subsidence. Therefore, the discrete delay terms  in our model act directly as the requisite MZ memory closure. These delays can be interpreted as phenomenological stand-ins for the thermodynamic inertia of the boundary layer, parameterizing the unresolved physics that dictate the visible 2D cloud field.

Crucially, the optimized delay scales natively align with the physics identified by the GWS analysis reported in Figs.~\ref{Fig_Dash_1to4}--\ref{Fig_Dash_5to8}. The slow delays parameterize the large-scale, slow-moving meteorological drivers. For instance, the intermediate delay ($\tau_{\mathrm{slw}_1}= 0.81$~h) provides the delayed feedback necessary to sustain mesoscale IGWs, while the longest delay ($\tau_{\mathrm{slw}_2} = 4.55$~h) encapsulates the overarching, slow-evolving capping effect of free-tropospheric subsidence and the gradual advection and pooling of specific humidity ($q$) required to precondition the environment for organized mesoscale cloud streets. Conversely, the fast delays parameterize rapid, highly localized thermodynamic adjustments. The shortest delay ($\tau_{\mathrm{fst}_1} = 0.16$~h, or $\sim 10$~minutes) captures the characteristic thermal turnover time ($t_*$), encapsulating the local perturbations of potential temperature ($\theta$) and the rapid phase changes ($\text{RH}$ saturation) associated with individual thermal updrafts and evaporative downdrafts. In this concatenated framework, the delays provide the vital memory information required to maintain the structural integrity of the turbulent wave-packets across the sequential simulated days.


\subsection*{Cloud Field Simulation: Phenomenological Analysis}
\noindent  The \textit{Supplementary Movie} presents an 80.8-hour out-of-sample stochastic simulation of the MSHDM in the physical domain (reconstructed via EOF multiplication), drawn from the ensemble analyzed in Figs.~\ref{Fig_Dash_1to4}--\ref{Fig_Dash_5to8}. The movie is also available here: \href{https://drive.google.com/file/d/1Kp8M-LtasOfQ1xxeP2LGYK_kKX5T7c3p/view?usp=drive_link}{https://drive.google.com/file/d/1Kp8M}.  

To effectively visualize the spatial organization and diurnal evolution of the simulated cloud fields, the reflectance anomalies are rendered using an asymmetrically red-grey-yellow colormap. Because clouds are highly reflective compared to the underlying continental surface, positive anomalies indicate regions with thicker or more continuous cloud cover than the temporal mean, while negative anomalies denote areas of comparatively reduced reflectance. Rather than imposing a binary ``cloud vs.~clear-sky'' threshold, this punctuated continuous color scale captures the full spectrum of the cloud field dynamics. Deep red identifies strong negative anomalies, highlighting regions of pronounced subsidence, severely dissipated clouds, or fully exposed land surface. The intermediate grey spectrum spans the near-zero baseline: dark grey indicates thinner-than-average or decaying clouds, while lighter grey represents the active, baseline shallow cumulus deck. Finally, the highest positive anomalies transition into bright yellow, acting as a distinct heat-map for the thickest, most energetic convective updraft cores. This targeted visualization strategy allows for immediate visual segregation between the actively growing convective wave-packets and the expanding regions of subsidence during the evening boundary layer collapse.

To assess the physical fidelity of the modeled AM-FM beat patterns and wave packets, we conduct a phenomenological analysis of the generated movie. Synthesizing the visual evidence from the asymmetrically scaled colormap, the most prominent feature across the integration is a highly stable, non-stationary cycle occurring at a roughly 12-hour interval with distinct convective peaks emerging around $t \approx 3, 15, 27, 39, 51, 63$, and $75$~hours. Within our concatenated daylight framework, this perfectly captures the asymmetric diurnal forcing of continental shallow cumulus. Each cycle begins with a rapid phase transition from dark grey (residual or decaying clouds) into widespread lighter grey, rapidly culminating in explosive patches of bright yellow. This sequence represents the breakdown of the boundary layer inversion and the onset of energetic convective updraft cores. Following each active phase (e.g., from $t=4.08$ to $t=9.08$~hours), a sharp, nonlinear collapse occurs; the bright yellow convective cores extinguish, and the domain is rapidly overtaken by deep red. This visually confirms the sharp temporal sub-harmonics (Pairs 2 and 3) acting to simulate severe late-afternoon boundary layer subsidence and cloud dissipation. Crucially, the model does not require an explicit ``sunrise'' forcing to restart. The emergence of the next day's convective wave is driven entirely by the parameterized delays ($\tau_{\mathrm{slw}}$ and $\tau_{\mathrm{fst}}$), demonstrating that these delays successfully encapsulates the thermodynamic inertia and moisture pooling from the previous diurnal cycle.

During the peak active phases of each cycle, the spatial field organizes into distinct, diagonal striations rather than chaotic noise. These alternating bands of bright yellow (active condensation and uplift) and red to dark grey (compensating subsidence and clearing) are the direct spatial manifestation of the $\sim 1.5$-hour Slow modes (Pairs 4 and 5). The emulator successfully translates the intermediate temporal delays ($\tau_{\mathrm{slw}_1}= 0.81$~h) into propagating mesoscale cloud streets. The convective overshooting of the yellow cores generates propagating IGWs, which in turn dictate the alternating strips of convective inhibition and triggered low-level lifting. A closer inspection of these dominant diagonal bands reveals a secondary layer of high-frequency spatial organization. The yellow striations are not continuous ribbons but are granular and highly textured, reflecting the Fast modes (Pairs 6 through 8, scaling from $\sim 60$ down to $\sim 40$~minutes).

As the simulation evolves, a distinct physical beat pattern emerges from the interference between the $\sim 1.5$-hour IGW wave packets and the $\sim 50$-minute fast convective lifecycles  (Table \ref{Table_Predicted_Freqs}). Within the broader mesoscale envelopes, localized bright yellow hotspots continuously flare up, mix, and collapse into darker grey. This dynamic represents the higher-frequency lifecycle of individual clouds across the  $40$-$60$-minute timescales (Table \ref{Table_Predicted_Freqs}).  

From a dynamical systems perspective, the most remarkable outcome of this extended visualization is the sustained phase coherence of the multiscale structure. Over seven concatenated semidiurnal cycles, the fine-scale fast modes remain tightly coupled to the mesoscale envelopes, generating a stable dynamics that natively produces subtle, structurally distinct spatial variability from day to day. The asymmetric diurnal collapse fully resets the spatial variance without damping the system into a homogeneous field, thoroughly validating the multi-scale, delay-driven hierarchical architecture of the MSHDM framework.


\subsection*{Iterative Calibration of Stochastic Forcing}
\noindent While the Bayesian optimization resolves the structural hyperparameters of the SDDE, the magnitude of the block-diagonal noise innovations ($\mathbf{\Sigma}_{\mathbf{s}}, \mathbf{\Sigma}_{\mathbf{x}}, \mathbf{\Sigma}_{\mathbf{y}}$) requires precise out-of-sample calibration to prevent energy collapse or artificial variance inflation. 

This requirement is deeply rooted in the principles of the Fluctuation-Dissipation Theorem from statistical mechanics \cite{majda2005information,Lucarini_Chekroun_PRL24}. In a stable stochastic system, the total variance is governed by a strict energetic balance: the continuous injection of energy via stochastic fluctuations must perfectly offset the energy lost to the system's deterministic dissipation (damping). Because the optimization strictly freezes the optimal deterministic operators ($\mathbf{A}$ and $\mathbf{\Phi}$), the system's out-of-sample dissipation rate is locked. Therefore, achieving the correct equilibrium variance requires fine-tuning the intensity of the stochastic driving forces.

To restore this fluctuation-dissipation balance, we implement an iterative calibration routine. The baseline noise covariance matrices are initialized directly from the empirical residuals of the in-sample regression stage. We then independently tune a multiplicative noise scaling vector, $\boldsymbol{\alpha} = [\alpha_{\mathbf{s}}, \alpha_{\mathbf{x}}, \alpha_{\mathbf{y}}]$, initially set to $\mathbf{1}$. During each iteration $k$, the baseline SDDE noise matrices are scaled by $\boldsymbol{\alpha}^{(k)}$, the model is integrated completely out-of-sample, and the resulting simulated block variances $V_{\text{sim}}^{(k)} = [\text{Var}(\mathbf{s}_{\text{sim}}^{(k)}), \text{Var}(\mathbf{x}_{\text{sim}}^{(k)}), \text{Var}(\mathbf{y}_{\text{sim}}^{(k)})]$ are evaluated against the target empirical observations $V_{\text{obs}}$.

Because the topological blocks ($\mathbf{s}, \mathbf{x}, \mathbf{y}$) are highly coupled, adjusting the noise in one layer cascades through the entire system. To ensure stable, monotonic convergence across the highly coupled manifolds, the scaling vector is updated via a damped ratio correction:
\begin{equation}
    \boldsymbol{\alpha}^{(k+1)} = \boldsymbol{\alpha}^{(k)} \circ \left( 1 + 0.5 \left( \sqrt{\frac{V_{\text{obs}}}{V_{\text{sim}}^{(k)}}} - 1 \right) \right),
\end{equation}
where $\circ$ denotes the Hadamard (element-wise) product. 

The square root directly reflects the fundamental stochastic relationship between the amplitude of the noise matrix (the standard deviation) and the integrated system variance. The $0.5$ damping factor serves as an under-relaxation parameter, critically preventing oscillatory overshooting caused by the cross-layer energy cascades. This iterative scaling process is repeated until the maximum block variance error converges to within $5\%$ of the observational targets. This final calibration step mathematically guarantees that the stochastic emulator produces ergodic trajectories that perfectly preserve the macroscopic energy envelope of the target system.

\medskip
\noindent{\bf DATA AVAILABILITY}

\noindent 
All satellite data used in this study are publicly available at \url{https://www.avl.class.noaa.gov/} (GOES-16 ABI level 1b radiances, \url{https://doi.org/10.7289/V5BV7DSR}, GOES-R Calibration Working Group and GOES-R Series Program, 2017).

\medskip
\noindent{\bf CODE AVAILABILITY}

\noindent Any relevant code necessary to reproduce the results presented here is available upon request.

\medskip
\noindent{\bf ACKNOWLEDGMENTS}

\noindent  
This work has been partially supported by the European Research Council (ERC) under the European Union’s Horizon 2020 research and innovation program (Grant Agreement 810370); by the Knell Family Institute for Artificial Intelligence, Weizmann Institute of Science; by the Planning and Budgeting Committee (VATAT) of the Council for Higher Education in Israel, as part of a grant for the climate research center titled "Integrating Climate Dynamics, Clouds, and Extreme Events through Teleconnections in Climate Networks;" by the Office of Naval Research (ONR) Multidisciplinary University Research Initiative (MURI) Grant N00014-20-1-2023, and by the National Science Foundation Grants DMS-2407483 and DMS-2407484.

\medskip
\noindent{\bf  AUTHOR CONTRIBUTIONS}

\noindent  MDC conceived the presented idea, led the theoretical analysis, the data-driven modeling, and the spatio-temporal analysis. IK led the cloud physics questions and interpretations. 
TD led the satellite data pipeline and its preprocessing, and led the EOF analysis.  HL led the numerical analysis of the DDE spectra. 
MDC, TD, HL, and IK discussed the results. MDC wrote the manuscript. All co-authors provided critical feedback and helped shape the research, analysis and manuscript.

\medskip
\noindent{\bf COMPETING INTERESTS}

\noindent  The authors declare no competing financial or non-financial interests.

}

\bibliographystyle{IEEEtran}
\bibliography{EMR_MSM_SDDEs_v25}

\newpage
\onecolumngrid

\end{document}